\documentstyle[12pt]{article}
\input amssym.def
\input amssym
\def \dst {\displaystyle}

\def \ul{\underline}
\title{Nonlinear Integrable Equations and Nonlinear Fourier Transform}
\author{A. S. Fokas, I. M. Gelfand, M. V. Zyskin}

\date{Preprint RU-95-09}

\begin{document}
\maketitle

\subsection*{Introduction}

In this paper we study nonlocal functionals whose kernels are homogeneous
generalized functions.  We also use such functionals to solve the Korteweg-de
Vries (KdV), the nonlinear Schr\"odinger (NLS) and the Davey-Stewartson
(DS) equations.

The solution of certain integrable equations in terms of formal power series
was obtained in  [4], [5].  In these
papers the solution  was expressed in a formal power series involving
scattering data.  In this paper in addition to developing techniques for
multiplying and inverting nonlocal functionals we also:

(a) Give the correct version of these series by giving meaning to the relevant
kernels , see (2.10) and (3.18)).

(b) We invert these
series to obtain scattering data in terms of initial data.

(c) Prove the convergence of these series.

(d) We extend these results to equations in two space dimensions.

\section{Nonlocal analytic functionals with\hfil\break homogeneous kernels}

The calculus of local functionals was developed by Gelfand and Dikii  [4].
Local functionals of one function $u(x)$ can be written
as multiple integrals using the kernels given by the $\delta$-function and
its derivatives. For example,
$$
\int u^2(x) dx = \int u(x_1) u(x_2) \delta(x_1 - x_2)\ dx_1 dx_2 \ ,
$$
$$
\int (u')^3 dx = \int u(x_1) u(x_2) u(x_3) \delta'(x - x_1)
\delta'(x - x_2) \delta'(x - x_3)\ dx dx_1 dx_2dx_3 \ ,
$$
$$
\begin{array}{r@{\;}l}
\dst\int u^2 (u')^3 dx =& \dst\int u(x_1) u(x_2) u(x_3) u(x_4) u(x_5)
\delta(x - x_1) \delta(x - x_2) \delta'(x - x_3)
\\ [12pt]
& \delta'(x - x_4) \delta'(x - x_5)\ dx dx_1 dx_2 \ldots dx_5 \ .
\end{array}
$$
and so on.

Nonlocal analytic functionals are those functionals whose kernels involve
homogeneous generalized functions.

In the case of a real variable a basis in the space of homogeneous
generalized functions is [1]

$$
\dst\frac{{x_+}^{\lambda - 1}}{\Gamma(\lambda)},\quad \frac{{x_-}^{\lambda -
1}}{\Gamma(\lambda)},\quad ( x + i0)^{\lambda}.
$$

In the case of a complex variable a basis in the space of homogeneous
generalized functions is
$$
\begin{array}{l}
z^s \bar{z}^{s + n} \hskip14pt n=0 , \pm 1, \pm 2, \ldots
\\ [12pt]
\delta-function \ \ and \ its \ derivatives
\end{array}
$$

{\bf Remarks.}

\noindent 1) \hskip12pt Only those functionals which make sense in the
framework of generalized functions are allowed. For example, the functionals
$$
\dst\int u(k_1) u(k_2) \frac{1}{k_1 + k_2 + i0} \delta(k_1 + k_2)\ dk_1
 dk_2,
$$
and
$$
\dst\int u(k_1) u(k_2) \frac{1}{k_1 + k_2 + i0} \frac{1}{k_1 + k_2 - i0}\ dk_1
 dk_2 ,
$$
are not allowed, while the functional
$$
\begin{array}{l}
\dst\int u^2 (k_1) u(k_2) \frac{1}{(k_1 + k_2)^2} dk_1 dk_2 :=
\frac{1}{2} \int^{\infty}_0 dk \int^{\infty}_{-\infty} dq
\frac{1}{k^2} \left( u^2 \left( \frac{q + k}{2} \right) u
\left( \frac{q - k}{2} \right) \right.
\\ [12pt]
\left.  + u^2 \left( \dst\frac{q - k}{2} \right) u \left(
\dst\frac{q + k}{2} \right) - 2 u^3(q) \right)
\end{array}
$$
is allowed. (see [1] for details).

\noindent 2) \hskip12pt  Local functionals are a particular case of
functionals with homogeneous kernels,for example,
$$
\int u^2(x) dx = \int u(x_1) u(x_2) \delta(x_1 - x_2)\ dx_1 dx_2 =
 \int u(x_1) u(x_2)  { \frac{{{(x_1-x_2)}_+}^{\lambda - 1}}{\Gamma(\lambda)}
\Biggr| }_{ \lambda = 0} dx_1 dx_2 .
$$

The product of two nonlocal analytic functionals is also a nonlocal analytic
functional whose kernel is the direct product
of the kernels of the two starting functionals. For example,
$$
\begin{array}{l}
\left( \dst\int u(x_1) u(x_2) \frac{{(x_1 -
x_2)}^{\lambda}}{\Gamma(\lambda + 1)} \right)
\cdot \left( \dst\int u(y_1) u(y_2) u(y_3)
\dst\frac{{(y_1 - y_2)}^{\mu_1}}{\Gamma(\mu_1 + 1)}
\frac{y_3^{\mu_2}}{\Gamma(\mu_2 + 1)} \right)
\\ [12pt]
= \dst\int u(x_1) u(x_2) u(x_3) u(x_4) u(x_5)
\frac{{(x_1 - x_2)}^{\lambda}}{\Gamma(\lambda + 1)}
\frac{{(x_3 - x_4)}^{\mu_1}}{\Gamma(\mu_1 + 1)}
\frac{{x_5}^{\mu_2}}{\Gamma(\mu_2 + 1)}
\ dx_1 \ldots dx_5
\end{array}
$$

There are certain relations in the algebra of nonlocal analytic
functionals.

\vspace{4pt}

{\bf Examples.}

1)
$$
\begin{array}{l}
\left( \dst\int u_1(x_1) u_2(x_2) \Theta(1 - x_1) \Theta(x_1 - x_2)
\Theta(x_2)\ dx_1 dx_2 \right)
\cdot
\\ [12pt]
\left( \dst\int v_1(y_1) v_2(y_2) v_3(y_3) \Theta(1 - y_1) \Theta(y_1 - y_2)
\Theta(y_2 - y_3) \Theta(y_3)\ dy_1 dy_2 dy_3 \right)
\\ [12pt]
= \dst\int^1_0 u_1(x_1) u_2(x_2) v_1(x_3) v_2(x_4) v_3(x_5) \Theta_{12}
\Theta_{34} \Theta_{45}\ dx_1 \ldots dx_5
\\ [12pt]
= \dst\int^1_0 u_1(x_1) u_2(x_2) v_1(x_3) v_2(x_4) v_3(x_5) \left( \Theta_{12}
\Theta_{23} \Theta_{34} \Theta_{45} +
\Theta_{13} \Theta_{32} \Theta_{24} \Theta_{45} \right.
\\ [12pt]
+ \Theta_{13} \Theta_{34} \Theta_{42} \Theta_{25}
+ \Theta_{13} \Theta_{34} \Theta_{45} \Theta_{52}
+ \Theta_{31} \Theta_{12} \Theta_{24} \Theta_{45}
\\ [12pt]
+ \Theta_{31} \Theta_{14} \Theta_{42} \Theta_{25}
+ \Theta_{31} \Theta_{14} \Theta_{45} \Theta_{52}
+ \Theta_{34} \Theta_{41} \Theta_{12} \Theta_{25}
\\ [12pt]
\left.
+ \Theta_{34} \Theta_{41} \Theta_{15} \Theta_{52}
+ \Theta_{34} \Theta_{45} \Theta_{51} \Theta_{12} \right)\ dx_1 \ldots dx_5
\end{array}
$$
where
$
\Theta(x) = \dst\frac{x^0_+}{\Gamma(1)} =
\left\{
\begin{array}{l}
1,\ x \geq 0\\
0,\ x < 0
\end{array}
\right.
, \mbox{ and } \Theta_{ij} := \Theta(x_i - x_j)
$

We have multiplied two functionals, with kernels of degree 0, and with
integration domain given by the  simplexes, $0 < x_1
< x_2 < 1$ and $0 < y_1 < y_2 < y_3 < 1$ ,respectively. For the product
functional, the integration domain is not a
simplex, but we can write it as sum of functionals, such that the integration
domain of each functional is a simplex . The simplexes are given by
all possible orderings of the letters $x_1, x_2, y_1, y_2, y_3$ such that
$x_1 < x_2$ and $y_1 < y_2 < y_3$ for all the orderings (all shuffles
of $(x_1 x_2)(y_1 y_2 y_3)$).

\vspace{4pt}

{\bf Remark.}
The functional $\int u_1(x_1) u_2(x_2) \Theta(y - x_1) \Theta(x_1 -
x_2) \Theta(x_2)\ dx_1 dx_2$ with $u_2(x) = \frac{1}{1 - x}$, $u_1(x)
= \frac{1}{x}$ is the dilogarithm $Li_2(y)$.

\vspace{4pt}

{\bf Example 2.}
$$
\begin{array}{l}
\dst\int u_1(x_1) u_2(x_2) u_3(x_3) u_4(x_4) u_5(x_5) \Theta_{12}
\Theta_{32} \Theta_{34} \Theta_{54}\ dx_1 \ldots dx_5
\\
- \left( \dst\int u_1(x_1) u_2(x_2) u_3(x_3) u_4(x_4) u_5(x_5)
\Theta_{12} \Theta_{32}\ dx_1 \ldots dx_3 \right)
\\
\left( \dst\int u_4(x_4) u_5(x_5)\right.
\Theta_{54} dx_4 = \dst\int u_1(x_1) u_2(x_2) \ldots u_5(x_5)
\Theta_{12} \Theta_{23} \Theta_{34} \Theta_{45}\ dx_1 \ldots dx_5
\end{array}
$$

\vspace{4pt}

{\bf Example 3.}
$$
\begin{array}{l}
\left( \dst\int u_1(k_1) u_2(k_2) \frac{1}{k_1 + i0} \delta(k_1 +
k_2) \frac{dk_1 dk_2}{(2\pi i)^2} \right)
\\ [12pt]
\cdot \left( \dst\int v_1(q_1) v_2(q_2) v_3(q_3)
\dst\frac{1}{(q_1 + i0)(q_1 + q_2 + i0)} \delta(q_1 +
q_2 + q_3) \cdot \dst\frac{dq_1 dq_2 dq_3}{(2 \pi i)^3} \right)
\\ [12pt]
= \dst\int u_1(k_1) u_2(k_2) v_1(k_3) v_2(k_4) v_3(k_5) \frac{\delta
(k_1 + k_2) \delta(k_1 + k_2 + k_3)}{(k_1 + i0)(k_3 + i0)(k_3 + k_4 +
i0)} \dst\frac{dk_1 \ldots dk_5}{(2\pi i)^5}
\\ [12pt]
= \dst\int u_1(k_1) u_2(k_2) v_1(k_3) v_2(k_4) v_3(k_5) \delta(k_1 +
k_2 + k_3 + k_4 + k_5)
\\ [12pt]
\cdot \left( p(1, 2, 3, 4) + p(1, 3, 2, 4) + p(1, 3, 4, 2) + p(1,
3, 4, 5) + p(3, 1, 2, 4) \right.
\\ [12pt]
\left. + p(3, 1, 4, 2) + p(3, 1, 4, 5) + p(3, 4, 1, 2) + p(3, 4, 1, 5) +
p(3, 4, 5, 1) \right)\ \dst\frac{dk_1 \ldots dk_5}{(2\pi i)^4}
\end{array}
$$
where
$$
\begin{array}{l}
p(m_1, m_2, m_3, m_4)
\\ [12pt]
= \dst\frac{1}{(k_{m_1} + i0)(k_{m_1} + k_{m_2} + i0)(k_{m_1} +
k_{m_2} + k_{m_3} + i0)(k_{m_1} + k_{m_2} + k_{m_3} + k_{m_4} + i0)}
\end{array}
$$

Nonlocal analytic functionals appear naturally in nonlinear
integrable equations. For example, in the $KdV$ equation the
transformation from the potential to scattering data and the inverse
transformation are given by nonlocal analytic functionals.

We will write these functionals using the inverse scattering formalism [3].
Alternatively,one could start directly from the nonlinear equation.

\section{Nonlocal functionals for the ${\bf KdV}$ equation}
\setcounter{equation}{0}

Let $u(x)$ be a $C^{\infty}$ real-valued function of a real variable
$x$, with the fast decrease as $x \rightarrow \pm \infty$.

We construct the following functionals of $u$ :
\begin{equation}
\begin{array}{l}
a(k) = 1 + \dst\sum^{\infty}_{n = 1} (-)^n \int
u(x_2) u(x_4) \ldots u(x_{2n }) \Theta_{12} \Theta_{23}
\Theta_{34} \cdot \ldots \cdot \Theta_{2n - 1, 2n }
\\ [12pt]
\quad \delta(x -x_1 + x_2 - \ldots + x_{2n}) \exp(2ikx)\ dx dx_1 dx_2
\ldots dx_{2n},
\end{array}
\end{equation}

\begin{equation}
\begin{array}{l}
b(k) = \dst\frac{1}{2i(k + i0)} \sum^{\infty}_{n = 0} (-)^{n + 1} \int
u(x_1) u(x_3) \cdot \ldots \cdot u(x_{2n + 1}) \Theta_{12} \Theta_{23}
\cdot \ldots \cdot \Theta_{2n, 2n + 1}
\\ [12pt]
\quad \delta(-x + x_1 - x_2 + x_3 - \ldots
+ x_{2n + 1}) \exp(-2ikx)\ dx dx_1 \ldots dx_{2n} ,
\end{array}
\end{equation}

\begin{equation}
\begin{array}{l}
\Psi(k, x) = 1 + \dst\sum^{\infty}_{n = 1} (-)^n \int
u(x_2) u(x_4) \cdot \ldots \cdot u(x_{2n}) \Theta(x_1 - x) \Theta_{21}
\Theta_{32} \cdot \ldots \cdot \Theta_{2n, 2n - 1}
\\ [12pt]
\quad \delta(x_0 - x_1 + x_2 - \ldots
+ x_{2n}) \exp(-2ikx_0)\ dx_0 dx_1 \ldots dx_{2n},
\end{array}
\end{equation}

\begin{equation}
\begin{array}{l}
\Phi(k, x) = 1 + \dst\sum^{\infty}_{n = 1} (-)^n \int
u(x_2) u(x_4) \cdot \ldots \cdot u(x_{2n}) \Theta(x - x_1)
\Theta_{12} \Theta_{23} \cdot \ldots \cdot \Theta_{2n - 1, 2n}
\\ [12pt]
\quad \delta(-x_0 - x_1 + x_2 - \ldots
+ x_{2n}) \exp(-2ikx_0)\ dx_0 dx_1 \ldots dx_{2n}.
\end{array}
\end{equation}

\vspace{4pt}

{\bf Remark.}
In the series, say, for $\Psi(k, x)$ we can integrate over $x_{2m +
1}$,\break $m = 0, 1, \ldots$, to get another formula for $\Psi(k,
x)$ :
$$
\begin{array}{l}
\Psi(k, x) = 1 + \dst\sum^{\infty}_{n = 1} \frac{(-)^n}{(2ik)^n}
\int u(x_1) u(x_2) \ldots u(x_{n - 1}) u(x_n)
\\ [12pt]
\quad \Theta(x_1 - x) \Theta_{21} \Theta_{32} \Theta_{43} \cdot
\ldots \cdot \Theta_{n, n - 1}
\\ [12pt]
\cdot (e^{-2ik(x - x_1)} - 1) (e^{-2ik(x_1 - x_2)} - 1) \ldots
(e^{-2ik(x_{n - 1} - x_n)} - 1)\ dx_1 \ldots dx_n .
\end{array}
$$

\vspace{2pt}

We define $S(k)$ by

\begin{equation}
S(k) = \frac{b(k)}{a(k)} .
\end{equation}

If $|1 - a(k)| < 1$, $S(k)$ can be written as
\begin{equation}
\begin{array}{l}
S(k) = - \dst\frac{1}{2i(k + i0)} \sum^{\infty}_{n = 0} \int u(x_1)
u(x_3) \ldots u(x_{2n + 1})
\\ [14pt]
\quad \Theta_{21} \Theta_{23} \Theta_{43} \Theta_{45} \ldots
\Theta_{2n, 2n - 1} \Theta_{2n, 2n + 1}
\\ [14pt]
\quad \delta(-x + x_1 - x_2 + x_3 - \ldots + x_{2n + 1}) \exp(-2ikx)
\ dx dx_1 dx_2 \ldots dx_{2n + 1}
\end{array}
\end{equation}

\vspace{2pt}
{\bf Convergence.}

The series (1)--(4) converge for $k \ne 0$. They also converge
at $k = 0$ if the moments of function $u(x)$, that is, $\int u(x)
dx$, $\int x u(x) dx$, $\int x^2 u(x) dx$, \ldots are small.
The series (5) is convergent if (1) and (2) are convergent, and
$|1 -a(k)| < 1$.

\vspace{6pt}

Indeed, let $k \ne 0$.Then
$$
\begin{array}{r@{\;}l}
|\Phi(k, x)|& \leq
1 + \dst\sum^{\infty}_{n = 1} \frac{1}{|k|^n} \int
|u(x_1) u(x_2) \ldots u(x_n)
\\ [14pt]
& \qquad \sin k(x - x_1) \sin k(x_1 - x_2) \ldots
\sin k(x_{n - 1} - x_n)
\\ [14pt]
& \qquad \Theta(x - x_1) \Theta_{12} \Theta_{23} \Theta_{34}
\cdot \ldots \cdot \Theta_{n - 1, n}\ dx_1 dx_2 \ldots dx_n
\\ [14pt]
& \leq 1 + \dst\sum^{\infty}_{n = 1} \frac{1}{n!} \frac{(\int|u(x)| dx)^n}{k^n}
\end{array}
$$

\vspace{3pt}
For all $k$, including $k = 0$,
$$
\begin{array}{r@{\;}l}
|\Phi(k, x)|& \leq
1 + \dst\sum^{\infty}_{n = 1} \int
|u(x_1) u(x_2) \ldots u(x_n) (x - x_1) (x_1 - x_2) \ldots (x_{n - 1} - x_n)|
\\ [14pt]
& \qquad \qquad \cdot \Theta(x - x_1) \Theta_{12} \cdot \ldots \cdot
\Theta_{n - 1, n}\ dx_1 dx_2 \ldots dx_n
\end{array}
$$

\vspace{3pt}
If the moments of function $u$ are small, the series is convergent
for $k = 0$ as well.

\vspace{3pt}
We will consider two operations on functionals (1)--(5): multiplication
and inversion.
These operations are infinite-dimensional analogues of multiplication
of functions and the inverse function.

There could be some relations among the products of functionals.

\vspace{4pt}

{\bf Example 1.}

Let us show that $a(k) a(-k) - b(k) b(-k) = 1$ for all $k\ne 0$ . Indeed,

$$
\begin{array}{l}
a(k) a(-k) = 1 + \dst\sum^{\infty}_{n = 1} \sum^n_{m = 0} (-)^n \int
u(x_2) u(x_4) \ldots u(x_{2n}) \Theta_{12} \Theta_{23} \ldots
\Theta_{2m - 1, 2m}
\\ [12pt]
\quad \cdot \Theta_{2m + 2, 2m + 1} \cdot \Theta_{2m + 3, 2m + 2}
\ldots \Theta_{2n, 2n - 1} e^{2ikx_0}
\\ [12pt]
\quad \delta(-x_0 + x_1 - x_2 + x_3 - x_4 + \ldots -x_{2n})\ dx_0 dx_1 dx_2
\ldots dx_{2n} ,
\end{array}
$$

$$
\begin{array}{l}
b(k) b(-k) = \dst\sum^{\infty}_{n = 1} \sum^{n - 1}_{m = 1} (-)^n \int
u(x_2) u(x_4) \ldots u(x_{2n}) \Theta_{12} \Theta_{23} \cdot \ldots
\cdot \Theta_{2m, 2m + 1}
\\ [12pt]
\quad \Theta_{2m + 3, 2m + 2} \cdot \Theta_{2m + 4, 2m + 3}
\cdot \ldots \cdot \Theta_{2n, 2n - 1} e^{2ikx_0}
\\ [12pt]
\quad \delta(-x_0 + x_1 - x_2 + x_3 - x_4 + \ldots -x_{2n})\ dx_0 dx_1 dx_2
\ldots dx_{2n} ,
\end{array}
$$

$$
\begin{array}{l}
a(k) a(-k) - b(k) b(-k) - 1 = \dst\sum^{\infty}_{n = 1} (-)^n \int
u(x_2) u(x_4) \ldots u(x_{2n})
\\ [12pt]
\exp(2ik(x_1 - x_2 + \ldots - x_{2n}))
\Theta_{23} \Theta_{34} \cdot \Theta_{45} \cdot \ldots \cdot
\Theta_{2n - 1, 2n}
\\ [12pt]
dx_1 \ldots dx_{2n} = 0,\quad k \ne 0.
\end{array}
$$

\vspace{10pt}
{}From ( ) one can see that $\overline{a(k)} = a(-k), \overline{b(k)} = b(-k)$;
therefore, $|a(k)| \geq 1$ and $|S(k)| = \left|
\dst\frac{b(k)}{a(k)} \right| = \dst\frac{|b(k)|}{\sqrt{1 +
|b(k)|^2}} \leq 1,\quad k \ne 0$ .

\vspace{4pt}

{\bf Example 2.}

Let us show that
$\Psi(-k, y) a(k) + \Psi(k, y) b(k) e^{2iky} =\Phi(k, y)$ . Indeed,

$$
\begin{array}{l}
\Psi(-k, y) a(k) = 1 + \dst\sum^{\infty}_{n = 1} (-)^n \int \left(
u(x_2) u(x_4) \ldots u(x_{2n}) \right)
\\ [12pt]
\quad \cdot \left( \Theta_{12} \Theta_{23} \Theta_{34} \cdot \ldots
\cdot \Theta_{2n - 1, 2n} \right.
\\ [12pt]
\quad +\dst\sum^{n - 1}_{m = 1} \left( \Theta(x_1 - y) \Theta_{21} \Theta_{32}
\cdot \ldots \cdot \Theta_{2m, 2m - 1} \Theta_{2m + 1, 2m + 2} \right.
\\ [12pt]
\quad \left. \left. \cdot \Theta_{2m + 2, 2m + 3} \cdot \Theta_{2n -
1, 2n} \right)  + \Theta(x_1 - y) \Theta_{21} \Theta_{32}
\cdot \ldots \cdot \Theta_{2n, 2n - 1} \right)
\\ [12pt]
\quad e^{-2ikx_0} \delta(x_0 + x_1 - x_2 + \ldots - x_{2n})\ dx_0
\ldots dx_{2n} ,
\end{array}
$$

also,

$$
\begin{array}{l}
\Psi(k, y) b(k) e^{2iky} = \dst\sum^{\infty}_{n = 1} (-)^{n + 1} \int
u(x_2) u(x_4) \cdot \ldots \cdot u(x_{2n})
\\ [12pt]
\quad \cdot \left( \Theta(x_1 - y) \Theta_{23} \Theta_{34}
\Theta_{45} \cdot \ldots \cdot \Theta_{2n - 1, 2n} \right.
\end{array}
$$
$$
\begin{array}{l}
\quad +\dst\sum^{n - 2}_{m = 1} \Theta(x_1 - y) \Theta_{21} \Theta_{32}
\cdot \ldots \cdot \Theta_{2m + 1, 2m} \cdot \Theta_{2m + 2, 2m + 3}
\\ [12pt]
\quad \left. \cdot \Theta_{2m + 3, 2m + 4} \cdot \ldots \cdot \Theta_{2n -
1, 2n} + \Theta(x_1 - y) \Theta_{21} \Theta_{32}
\cdot \ldots \cdot \Theta_{2n - 1, 2n} \right)
\\ [12pt]
\quad e^{-2ikx_0} \delta(x_0 + x_1 - x_2 + \ldots - x_{2n})\ dx_0
dx_1 \ldots dx_{2n} ,
\end{array}
$$
and
$$
\begin{array}{l}
\Psi(-k, y) a(k) + \Psi(k, y) b(k) e^{2iky} = 1 +
\dst\sum^{\infty}_{n = 1} (-)^n
\int u(x_2) u(x_4) \cdot \ldots \cdot u(x_{2n})
\\ [12pt]
\quad \Theta(y - x_1) \Theta_{12} \Theta_{23}
\cdot \ldots \cdot \Theta_{2n - 1, 2n} e^{-2ikx_0}
\\ [12pt]
\quad \delta(x + x_1 - x_2 - \ldots + x_{2n - 1} - x_{2n})\ dx_0
dx_1 \ldots dx_{2n} = \Phi(k, y).
\end{array}
$$

\vspace{4pt}

{\bf Example 3.}

Let us take $ S(k)$ ,given by the formal series (6), and $a(k)$ , $b(k)$ ,
given by the convergent series (1), (2). We can prove the following relation
for the series in {u(x)}:

$$
a(k) S(k) = b(k).
$$

\noindent The computation is similar to that of the Examples 1 and 2.

\ul{\bf Inversion}.

Let us define $S(x) = \dst\int S(k) e^{2ikx} \frac{dk}{\pi}$. We suppose
that $S(x)$ is a fast decreasing function as $x \rightarrow + \infty$. Formula
(6) can be written as
$$
\begin{array}{l}
\dst\frac{d}{dx} S(x) = -\left( u(x) + \dst\int u(x_1) u(x_3) \Theta_{21}
\Theta_{23} \delta(x - x_1 + x_2 - x_3) dx_1 dx_2 dx_3 \right.
\\ [12pt]
\quad + \ldots \int u(x_1) u(x_3) \ldots u(x_{2n + 1}) \Theta_{21} \Theta_{23}
\Theta_{43} \Theta_{45} \ldots \Theta_{2n, 2n - 1} \Theta_{2n, 2n +
1}
\\ [12pt]
\quad \Bigl. \delta(x - x_1 + x_2 - x_3 + \ldots - x_4)\ dx_1 dx_2
\ldots dx_{2n + 1} + \ldots \Bigr)
\end{array}
$$

\noindent (here the right-hand side is a formal series in $u(x)$ ).

We can invert it, that is, we can express $u(x)$ in terms of $S(x)$ by the
formal
series $u(x) = \dst\sum^{\infty}_{k = 1} W_k(x)$, where $W_k(x)$ is a nonlocal
analytic functional of degree $k$ in $S(x)$ ($S(x)$ has degree one). The
functionals $W_k(x)$ are determined recursively:

$$
\begin{array}{r@{\;}l}
W_1(x) =& -\dst\frac{d}{dx} S(x) ,
\\ [12pt]
W_2(x) =& -\dst\int \frac{d}{dx_1} S(x_1) \frac{d}{dx_3} S(x_3)
\Theta_{21} \Theta_{23} \delta(x - x_1 + x_2 - x_3)\ dx_1 dx_2 dx_3
\\ [12pt]
=& -\dst\int S(x_1) S(x_3) \left( \delta(x_2 - x_1) \delta(x_2 - x_3)
\delta(x - x_1 + x_2 - x_3) \right.
\\ [12pt]
&\hspace{6.5pc} + \delta(x_2 - x_1) \Theta_{23} \delta'(x - x_1 + x_2 -
x_3)
\\ [12pt]
&\hspace{6.5pc} + \Theta_{21} \delta(x_2 - x_3) \delta'(x - x_1 + x_2 -
x_3)
\\ [12pt]
&\hspace{6.5pc} \left. + \Theta_{23} \Theta_{23} \delta''(x - x_1 + x_2 -
x_3) \right)\ dx_1 dx_2 dx_3
\end{array}
$$
$$
\begin{array}{r@{\;}l}
\phantom{W_1(x)} =& -\dst\int S(x_1) S(x_3) \delta(x_2 - x_1) \delta(x_2 - x_3)
\delta(x - x_1 + x_2 - x_3)\ dx_1 dx_2 dx_3
\\ [12pt]
=& -S^2(x) = \dst\frac{d}{dx} \int S(x_1) S(x_2) \Theta(x_1 - x)
\delta(x_1 - x_2)\ dx_1 dx_2 ,
\end{array}
$$
(we integrated by parts),
$$
\begin{array}{l}
W_n(x) = -\dst\sum^{n }_{k = 2}
\sum^{}_{{m_1, m_2 \ldots m_k \geq 1\quad\ }\atop {m_1 +
m_2 + \ldots + m_k = n}} \int W_{m_1}(x_1)
W_{m_2}(x_3) \ldots W_{m_k}(x_{2k + 1})
\\ [12pt]
\quad \Theta_{21} \Theta_{23} \cdot \Theta_{43} \Theta_{45} \cdot \ldots
\cdot \Theta_{2k, 2k - 1} \Theta_{2k, 2k + 1}
\\ [12pt]
\quad \delta(x - x_1 + x_2 - \ldots
+ x_{2k} - x_{2k + 1})\ dx_1 dx_2 \ldots dx_{2k + 1} .
\end{array}
$$

{\bf Lemma}.
\begin{em}
Consider the functionals of $S(x),$ given by the formal series
\begin{equation}
\begin{array}{l}
\tilde{u}(x) = -\dst\frac{d}{dx} S(x) + \frac{d}{dx} \int S(x_1)
S(x_2) \Theta(x_1 - x) \delta(x_1 - x_2)\ dx_1 dx_2
\\ [12pt]
-\dst\frac{d}{dx} \sum^{\infty}_{n = 1} \int S(x_1) S(x_2) \ldots
S(x_{2n + 1}) \Theta(-x_1 + x_2)
\\ [12pt]
\quad \Theta(-x_1 + x_2 - x_3 + x_4) \ldots \Theta(-x_1 + x_2 -
\ldots + x_{2n}) \Theta(-x_{2n + 1} + x_{2n})
\\ [12pt]
\quad \Theta(-x_{2n + 1} + x_{2n} - x_{2n - 1}
+ x_{2n - 2}) \ldots \Theta(-x_{2n + 1} + x_{2n} - \ldots + x_2)
\\ [12pt]
\quad \delta(x - x_1 + x_2 - x_3 + \ldots - x_{2n + 1})\ dx_1 dx_2 \ldots
dx_{2n + 1}
\end{array}
\end{equation}
$$
\begin{array}{l}
+ \dst\frac{d}{dx} \sum^{\infty}_{n = 2} \int S(x_1) S(x_2) \ldots
S(x_{2n})
\\ [12pt]
\quad \Theta(-x_1 + x_2) \Theta(-x_1 + x_2 - x_3 + x_4) \ldots \Theta(-x_1
+ x_2 - \ldots + x_{2n - 2})
\\ [12pt]
\quad \Theta(x_1 - x) \Theta(x_1 - x_2 + x_3 - x) \ldots \Theta(x_1 -
x_2 + x_3 - \ldots + x_{2n - 1} - x)
\\ [12pt]
\quad \delta(x - x_2 + x_3 - x_4 + \ldots - x_{2n})\ dx_1 \ldots
dx_{2n}.
\end{array}
$$

\vspace{4pt}

\begin{equation}
\begin{array}{l}
\tilde{\Psi}(x, y) = \delta(x) - \dst\sum^{\infty}_{n = 0} \Theta(-x)
\int S(x_1) S(x_3) \ldots S(x_{2n + 1})
\\ [12pt]
\quad \Theta(-x_1 + x_2) \Theta(-x_1 + x_2 - x_3 + x_4) \ldots
\Theta(-x_1 + x_2 - \ldots + x_{2n})
\\ [12pt]
\quad \Theta(x_1 - y) \Theta(x_1 - x_2 + x_3 - y) \ldots
\Theta(x_1 - x_2 + x_3 - \ldots + x_{2n + 1} - y)
\\ [12pt]
\quad \delta(x - y + x_1 - x_2 + x_3 - \ldots + x_{2n + 1})\ dx_1
\ldots dx_{2n + 1}
\end{array}
\end{equation}
$$
\begin{array}{l}
+ \dst\sum^{\infty}_{n = 1} \Theta(-x) \int S(x_1) S(x_2) \ldots
S(x_{2n})
\\ [12pt]
\quad \Theta(-x_1 + x_2) \Theta(-x_1 + x_2 - x_3 + x_4) \Theta(-x_1 + x_2 -
\ldots + x_{2n - 2})
\\ [12pt]
\quad \Theta(x_1 - y) \Theta(x_1 - x_2 + x_3 - y) \ldots \Theta(x_1 -
x_2 + x_3 - \ldots + x_{2n - 1} - y)
\\ [12pt]
\quad \delta(x - x_1 + x_2 - \ldots + x_{2n})\ dx_1 dx_2 \ldots
dx_{2n} .
\end{array}
$$

\vspace{6pt}

We can substitute in these series $S(x)$ as a functional of $\{ u(x)\}$ ,
given by the formal series

\begin{equation}
\begin{array}{l}
S(x) = \dst\int \Theta(x_1 - x) u(x_1) dx_1 + \dst\sum^{\infty}_{n =
1} \int u(x_1) u(x_3) \ldots u(x_{2n + 1})
\\ [12pt]
\Theta_{21} \Theta_{23} \Theta_{43} \Theta_{45} \cdot \ldots \cdot
\Theta_{2n, 2n - 1} \Theta_{2n, 2n + 1}
\\ [12pt]
\cdot \delta(x_0 - x_1 + x_2 - x_3 + \ldots - x_{2n + 1})\ dx_0
dx_1 \ldots dx_{2n + 1}.
\end{array}
\end{equation}

\vspace{2pt}
As a result of this substitution, we will have $\tilde{u}(x)$ and
$\tilde{\Psi}(x, y)$ given by formal series in $\{u(x)\}$. Moreover,

$\tilde{u}(x) = u(x)$ ,

$
\tilde{\Psi}(x, y) = \Psi(x, y) := \dst\int \Psi(k, y) e^{2ikx}
\frac{dk}{\pi}$, where $\Psi(k, y)$ is given by $(3)$.
\end{em}

\vspace{6pt}

{\bf Proof}

We will prove the lemma by induction in the degree of $\{ u(x) \}$.

\vspace{2pt}
1) In the first order in $\{ u(x) \}$
$$
\begin{array}{l}
S_{(1)}(x) = \dst\int \Theta(x_1 - x) u(x_1)\ dx_1 ,
\\ [12pt]
\tilde{u}(x)_1 = -\dst\frac{d}{dx} S_{(1)}(x) = u(x) ,
\\ [12pt]
\tilde{\Psi}_{(1)} (x, y) = -\Theta(-x) \dst\int S_{(1)} (x_1)
\delta(x - y + x_1)\ dx_1
\\ [12pt]
\quad = -\Theta(-x) \dst\int \Theta(x_2 - x_1) u(x_2) \delta(x - y +
x_1)\ dx_1 dx_2
\\ [12pt]
\quad = -\dst\int \Theta(x_1 - y) \Theta(x_2 - x_1) u(x_2) \delta(x - y +
x_1)\ dx_1 dx_2
\\ [12pt]
\quad -\dst\int \Theta(x_1 - y) \Theta(x_2 - x_1) u(x_2) \delta(x - x_1 +
x_2)
\\ [12pt]
\quad \cdot dx_1 dx_2 = \Psi_{(1)} (x, y) .
\end{array}
$$
\vspace{2pt}
(In the last step we have made the change of variables $x_1 \rightarrow y +
x_2 - x_1$).

2) Suppose that we have proved that
$$
\begin{array}{l}
\tilde{\Psi}(x, y) = \delta(x) + \dst\sum^N_{n = 1} (-)^n \int u(x_2)
u(x_4) \ldots u(x_{2n}) \Theta(x_1 - y) \Theta_{21} \Theta_{32}
\ldots \Theta_{2n, 2n - 1}
\\ [12pt]
\quad \delta(x - x_1 + x_2 - \ldots + x_{2n})\ dx_1 dx_2 \ldots
dx_{2n} + O(u^{N + 1}) .
\end{array}
$$
{}From the definition of $\tilde{\Psi}(x, y)$
$$
\tilde{\Psi}(x, y) = \delta(x) - \Theta(-x) \int \tilde{\Psi}(x_1, y)
S(y - x - x_1)\ dx.
$$
But $S(x)$ is a series in $\{ u(x) \}$ with terms of degree $\geq 1$
in $u$, therefore, if we know $\tilde{\Psi}(x, y)$ as functionals of
$\{ u(x) \}$ up to degree $n$, we can compute it in the next order
$(n + 1)$.

Notice that
$$
\begin{array}{l}
(-)^n \dst\int u(x_2) u(x_4) \ldots u(x_{2n)} \Theta(x_1 - y)
\Theta_{21} \Theta_{32} \Theta_{43} \ldots \Theta_{2n, 2n - 1}
\\ [12pt]
\quad \delta(x - x_1 + x_2 - \ldots x_{2n})\ dx_1 dx_2 \ldots + dx_{2n}
\\ [12pt]
= (-)^n \dst\int u(x_2) u(x_4) \ldots u(x_{2n}) \Theta(x_1 - y)
\Theta_{21} \Theta_{32} \Theta_{43} \ldots \Theta_{2n, 2n - 1}
\\ [12pt]
\quad \delta(x - y + x_1 - x_2 + x_3 -\ldots + x_{2n - 1})\ dx_1 dx_2 \ldots
dx_{2n}
\\ [12pt]
= (-)^{n + 1} \dst\frac{d}{dx} \int u(x_2) u(x_4) \ldots u(x_{2n})
\Theta(x_1 - y) \Theta_{21} \Theta_{32} \Theta_{43} \ldots
\Theta_{2n, 2n - 1} \Theta_{2n + 1, 2n}
\\ [12pt]
\cdot \delta \left( x - y + x_1 - x_2 + x_3 - \ldots + x_{2n - 1} - x_{2n} +
x_{2n + 1} \right)\ dx_1 dx_2 dx_3 \ldots dx_{2n} dx_{2n + 1}
\end{array}
$$
(In the first step we have used the change of variables $x_1 \rightarrow
x_2 + y - x_1, x_3 \rightarrow x_2 + x_4 - x_3 , \ldots , x_{2n - 1}
\rightarrow
x_{2n - 2} + x_{2n} - x_{2n - 1}$).

Also, $\delta(x) = -\dst\frac{d}{dx} \int \Theta(x_1 - y) \delta(x - y + x_1)
dx_1$, and
$$
\begin{array}{l}
\tilde{\Psi}_{N + 1}(x, y) = -\dst\sum^N_{m = 0} \Theta(-x)
\int \tilde{\Psi}_{(m)}(x_0, y) S_{(N + 1 - m)}(y - x - x_0)\ dx_0
\\ [12pt]
= -\dst\sum^N_{m = 0} \Theta(-x) (-)^{m + 1} \dst\int \frac{d}{dx_0}
\left( u(x_2) u(x_4) \ldots u(x_{2m}) \right.
\\ [12pt]
\Theta(x, y_1) \Theta_{21}
\Theta_{32} \Theta_{43} \ldots \Theta_{2m, 2m - 1} \Theta_{2m + 1, 2m}
\\ [12pt]
\cdot \delta \left( \left. x_0 - y + x_1 - x_2 + x_3 - \ldots + x_{2m
+ 1} \right) \right)
\\ [12pt]
\cdot S_{(N + 1 - m)} (y - x - x_0) dx_0 dx_1 dx_2 \ldots dx_{2m + 1}
\\ [12pt]
 = \Theta(-x) \dst\int u(x_2) u(x_4) \ldots u(x_{2n + 2})
\end{array}
$$
$$
\begin{array}{l}
\left( \dst\sum^N_{m = 0} (-)^{m + 1} \Theta(x_1 - y) \Theta_{21} \Theta_{32}
\ldots \Theta_{2m, 2m - 1} \Theta_{2m + 1, 2m} \right.
\cdot \left( \Theta_{2m + 2, 2m + 1} + \Theta_{2m + 1, 2m + 2} \right)
\\ [12pt]
\left. \phantom{\dst\sum} \left( \Theta_{2m + 3, 2m + 2} \Theta_{2m +
3, 2m + 4} \cdot \ldots \cdot \Theta_{2N + 1, 2N} \Theta_{2N + 1, 2N
+ 2} \right) \right)
\\ [12pt]
\cdot \delta(-x + x_1 - x_2 + x_3 - \ldots + x_{2N + 1} - x_{2N + 2})
\ dx_1 dx_2 \ldots dx_{2N + 2}
\\ [12pt]
= (-)^{N + 1} \Theta(-x) \dst\int u(x_2) u(x_4) \ldots u(x_{2N + 2})
\\ [12pt]
\left( \Theta(x_1 - y_1) \Theta_{21} \Theta_{32} \Theta_{43} \cdot \ldots
\cdot \Theta_{2N + 2, 2N + 1} -(-)^{N + 1} \right.
\\ [12pt]
\cdot \left. \Theta(x_1 - y) \Theta_{12} \Theta_{32} \Theta_{34}
\Theta_{54} \cdot \ldots \cdot \Theta_{2N + 1, 2N} \Theta_{2N + 1, 2N
+ 2} \right)
\\ [12pt]
\delta(-x + x_1 - x_2 + x_3 - \ldots + x_{2N + 1} - x_{2N + 2})
\ dx_1 dx_2 \ldots x_{2N + 2}
\end{array}
$$

The second term in the last expression  is zero, because the volume of the
integration domain vanishes ; the first term coincides with the term of
degree $(N + 1)$ in $\Psi(x, y)$, see (3).

To prove the formula for $\tilde{u}(x)$ we use the relation
$\tilde{u}(x) = -\dst\frac{d}{dx} \int\break S(x - x_1) \tilde{\Psi}(x_1, x)
dx_1$, which follows from the definition of $\tilde{u}$ and
$\tilde{\Psi}$. We know both $S(x)$ and $\tilde{\Psi}(x, y)$ as
functionals in $\{ u(x) \}$. The calculation of the same type as
above gives that only the first order term in $\{ u(x) \}$ is not zero:
$$
\begin{array}{l}
\tilde{u}(x) = \dst\frac{d}{dx} \dst\sum^{\infty}_{n = 1} (-)^n \int
u(x_2) u(x_4) \ldots u(x_{2n}) \delta(-x_1 + x_2 - \ldots + x_{2n})
\\ [12pt]
\cdot \left( \Theta(x_1 - x) \Theta_{21} \Theta_{32} \Theta_{43} \cdot \ldots
\cdot \Theta_{2n, 2n - 1} -(-)^n \right.
\\ [12pt]
\left. \Theta(x_1 - x) \Theta_{12} \Theta_{32} \Theta_{34}
\Theta_{54} \cdot \ldots \cdot \Theta_{2n - 1, 2n - 2} \Theta_{2n -
1, 2n} \right)
\\ [12pt]
dx_1 dx_2 \ldots dx_{2n} = u(x)
\end{array}
$$

The formula (7) can be written as follows:
\begin{equation}
\begin{array}{l}
u(x) = 4 \dst\sum^{\infty}_{n = 1} \int \frac{S(k_1) S(k_2) \ldots
S(k_n)(k_1 + k_2 + \ldots + k_n)}{(k_1 + k_2 + i0)(k_2 + k_3 + i0)
\ldots (k_{n - 1} + k_n + i0)}
\\ [12pt]
\exp(2ikx) \delta(k - k_1 - k_2 - \ldots - k_n)\ \dst\frac{dk dk_1 dk_2
\ldots dk_n}{(2\pi i)^n}
\end{array}
\end{equation}

Any polynomial in $u(x)$ and its derivative can be written as the
functional of the same type, but with some polynomial in $\{ k_i\}$
in the numerator. This property is similar to the usual Fourier transform
of the linear function of $u$ and its derivatives .
Therefore it is natural to call transformation (10) the Nonlinear
Fourier transform of $u$ :
\begin{equation}
\begin{array}{l}
u^{(d_1)} u^{(d_2)} \ldots u^{(d_m)} = 4^m(2i)^{d_1 + d_2 + \ldots +
d_m}
\\ [12pt]
\dst\sum^{\infty}_{n = m} \int \frac{S(k_1) S(k_2) \ldots
S(k_n)}{(k_1 + k_2 + i0)(k_2 + k_3 + i0) \ldots (k_{n - 1} + k_n +
i0)}
\\ [12pt]
\exp(2ikx) \cdot \delta(k - k_1 - k_2 - \ldots - k_n) \cdot \mbox{Sym}
_{d_1, d_2 \ldots d_m}
\\ [12pt]
\dst\sum_{1 \leq p_1 < p_2 < \ldots < p_{m -
1} < n} (k_1 + k_2 + \ldots + k_{p_1})^{d_1 + 1}(k_{p_1} + k_{p_1 +
1})
\\ [12pt]
(k_{p_1 + 1} + k_{p_1 + 2} + \ldots +
k_{p_2})^{d_2 + 1} (k_{p_2} + k_{p_2 + 1}) (k_{p_2 + 1} + k_{p_2 + 2}
+ \ldots + k_{p_3})^{d_3 + 1}
\\ [12pt]
\cdot (k_{p_3} + k_{p_3 + 1}) (k_{p_{m - 1}+1} + k_{p_{m - 1}+2} + \ldots +
k_n)^{d_m + 1} \dst\frac{dk dk_1 dk_2 \ldots dk_n}{{(2\pi i)}^n} .
\end{array}
\end{equation}

{\bf Examples.}
\begin{equation}
\begin{array}{l}
6u u_x = 32i \dst\sum^{\infty}_{n = 2} \int \frac{S(k_1) S(k_2) S(k_3) \ldots
S(k_n)}{(k_1 + k_2 + i0)(k_2 + k_3 + i0) \ldots (k_{n - 1} + k_n +
i0)}
\\ [12pt]
\exp(2ikx) \cdot \delta(k - k_1 - k_2 - \ldots - k_n)
\\ [12pt]
\cdot k \cdot \left( (k_1 + k_2 + \ldots + k_n)^3 - ({k_1}^3 +
{k_2}^3 + \ldots + {k_n}^3) \right)\ \dst\frac{dk\ dk_1\ dk_2 \ldots
dk_n}{(2\pi i)^n}
\\ [12pt]
u_{xxx} + 6u u_x = -32i \dst\sum^{\infty}_{n = 1} \int \frac{S(k_1)
S(k_2) S(k_3) \ldots S(k_n)}{(k_1 + k_2 + i0)(k_2 + k_3 + i0) \ldots
(k_{n - 1} + k_n + i0)}
\\ [12pt]
\exp(2ikx) \cdot \delta(k - k_1 - k_2 - \ldots - k_n)
\\ [12pt]
\cdot k \cdot \left( {k_1}^3 + {k_2}^3 + \ldots + {k_n}^3) \right)\
\dst\frac{dk dk_1 dk_2 \ldots dk_n}{(2\pi i)^n}
\end{array}
\end{equation}

\vspace{6pt}
We see that for the differential polynomial $u_{xxx} + 6u u_x$ the
corresponding
polynomial in $\{ k \}$ after the nonlinear Fourier transform is $({k_1}^3 +
{k_2}^3 + \ldots + {k_n}^3 )$, n= 1,2, \ldots .

We can use this fact to solve some nonlinear equation. Let $S(k, t) = S_{0}(k)
e^{8ik^3t}$, and $u(x, t)$ is defined
as a functional of $S(k, t)$ by (10) (with $S(k,t)$ instead of $S(k)$ ). Such
$u(x,t)$  solves the $KdV$ equation

$$
-\dst\frac{d}{dt} u(x, t) = \frac{\partial^3}{\partial x^3} u(x, t) + 6u
\frac{\partial u}{\partial x}.
$$
\noindent Indeed, for such $u(x,t)$ the right-hand side of the equation is
expressed in terms of $S(k,t)$ by (12), and it coincides with
$-\dst\frac{d}{dt} u(x, t)$.

Let us find polynomials in $u$ and derivatives of $u$ such that after the
Nonlinear Fourier Transform the corresponding polynomial in ${k}$ is given
by $\dst\sum_{i} {k_i}^{2l - 1 }$ , $l = 1,2, \ldots $ .In order to do this,
let
us first compute the resolvent: 
$$
\begin{array}{l}
\Psi(k, x) = 1 + \dst\sum^{\infty}_{n = 1} \int \frac{S(k_1) S(k_2)
S(k_3) \ldots S(k_n)}{(k + k_1 + i0)(k_2 + k_3 + i0) \ldots (k_{n -
1} + k_n + i0)}
\\ [6pt]
\exp 2i(k_1 + \ldots + k_n)x\ \dst\frac{dk_1 \ldots
dk_n}{(2\pi i)^n} ,
\end{array}
$$

\vspace{6pt}

$$
\begin{array}{l}
R(k, x) = \Psi(k, x) \Psi(-k, x - 0)
\\ [12pt]
= 1 + \dst\sum^{\infty}_{n = 1} \int S(k_1) S(k_2) \ldots S(k_n)
\exp(2i(k_1 + \ldots + k_n)x)
\cdot \left( {\dst\sum^{n - 1}_{m = 1}} \right.
\\ [12pt]
\frac{1}{(k_1 + k_2 + i0)(k_2 +
k_3 + i0) \ldots (k_{m - 1} + k_m + i0)(k_m + k + i0)(k_{m + 1} - k +
i0)(k_{m + 1} + k_{m + 2} + i0) \ldots (k_{n - 1} + k_n + i0)}
\\ [12pt]
+ \dst\frac{1}{(k + k_1 + i0)(k_1 + k_2 + i0) \ldots (k_{n -
1} + k_n + i0)}
\\ [12pt]
+ \left. \dst\frac{1}{(-k + k_1 + i0)(k_1 + k_2 + i0) \ldots (k_{n
- 1} + k_n + i0)} \right) \dst\frac{dk_1 \ldots dk_n}{(2\pi i)^n}
\\ [12pt]
= 1 + \dst\sum^{\infty}_{n = 1} \int \frac{S(k_1) S(k_2) \ldots
S(k_n) \exp(2i(k_1 + \ldots k_n)x)}{(k_1 + k_2 + i0)(k_2 + k_3 + i0)
\ldots (k_{n - 1} + k_n + i0)}
\\ [12pt]
\left( \dst\sum^{n - 1}_{m = 1} \left( \frac{1}{k_m + k + i0} +
\frac{1}{k_{m + 1} - k + i0}  \right) + \frac{1}{k + k_1 + i0} +
\frac{1}{-k + k_1 + i0} \right) \dst\frac{dk_1 \ldots dk_n}{(2\pi i)^n}
\\ [12pt]
\sim 1 - 2\dst\sum^{\infty}_{n = 1} \int \frac{S(k_1) S(k_2) \ldots
S(k_n)}{(k_1 + k_2 + i0)(k_2 + k_3 + i0) \ldots (k_{n - 1} + k_n +
i0)} \exp 2i(k_1 + \ldots + k_n)x
\\ [12pt]
\dst\sum^{\infty}_{l = 1} \frac{1}{k^{2l}} (k^{2l - 1}_1 +
k^{2l - 1}_2 + \ldots + k^{2l - 1}_n) \cdot \frac{dk_1 \ldots
dk_n}{(2\pi i)^n}.
\end{array}
$$

\vspace{4pt}

{\bf Lemma}.

\begin{em}
$1)$
$$
\begin{array}{r@{\;}l}
I_1 =&  \dst\frac{1}{4} \dst\int\ u(x) dx
\\ [12pt]
& \quad = \dst\sum^{\infty}_{n = 1} \int \frac{S(k_1) S(k_2) \ldots
S(k_n) k\exp(2ikx)}{(k_1 + k_2 + i0)(k_2 + k_3 + i0) \ldots (k_{n -
1} + k_n + i0)}
\\ [12pt]
& \quad \delta(-k + k_1 + \ldots + k_n)\ \dst\frac{dk dk_1 \ldots dk_n}{(2\pi
i)^n}\ dx
\end{array}
$$
$$
\begin{array}{r@{\;}l}
I_3 =& - \dst\frac{1}{16} \dst\int (3u^2 + u'')
\\ [12pt]
& \quad = \pi \dst\sum^{\infty}_{n = 1} \dst\int \frac{S(k_1) \ldots
S(k_n)({k_1}^3 + {k_2}^3 + \ldots + {k_n}^3)}{(k_1 + k_2 + i0)(k_2 +
k_3 + i0) \ldots (k_{n - 1} + k_n + i0)}
\\ [12pt]
& \quad \delta(k_1 + \ldots + k_n)\ \dst\frac{dk_1 \ldots dk_n}{(2\pi
i)^{n}}
\\ [12pt]
I_5 =& \dst\frac{1}{64} \dst\int ( 10u^3 + 10u u'' + 5(u')^2 + u^{1v})
\\ [12pt]
& \quad = \pi \dst\sum^{\infty}_{n = 1} \int \frac{S(k_1) \ldots
S(k_n) ({k_1}^5 + {k_2}^5 + \ldots + {k_n}^5)}{(k_1 + k_2 + i0)(k_2 +
k_3 + i0) \ldots (k_{n - 1} + k_n + i0)}
\\ [12pt]
& \quad \delta(k_1 + \ldots + k_n)\ \dst\frac{dk_1 \ldots dk_n}{(2\pi
i)^{n}}
\end{array}
$$
\vspace{8pt}
\centerline \ldots
\vspace{8pt}

$2)$ let $S(k, t) = S(k) e^{ik^{2l + 1}t}, l = 0, 1, 2, \ldots,$ and
$$
\begin{array}{l}
I_{l, m} = -\dst\frac{1}{2} \sum^{\infty}_{n = 1} \int
\frac{S(k_1, t) S(k_2, t) \ldots S(k_n, t)}{(k_1 + k_2 + i0)(k_2 +
k_3 + i0) \ldots (k_{n - 1} + k_n + i0)}
\\ [12pt]
({k_1}^{2m + 1} + {k_2}^{2m + 1} + \ldots + {k_n}^{2m + 1})
\delta(k_1 + \ldots + k_n)\ \dst\frac{dk_1 \ldots dk_n}{(2\pi i)^{n - 1}}
\end{array}
$$

Such functionals $I_{l, m}$ are conserved$:$ $\dst\frac{d}{dt} I_{l, m} = 0$.
\end{em}

\vspace{6pt}
{\bf Theorem 2.1 (Solution of the $ KdV $ equation).}

\vspace{2pt}
Consider the Cauchy problem for the $KdV$ equation

\begin{equation}
-\frac{\partial}{\partial t} u(x, t) = \frac{\partial^3}{\partial
x^3} u(x, t) + 6u \frac{\partial u}{\partial x}  , t\geq 0 \quad u(x, 0) = u(x)
\end{equation}

\vspace{4pt}

\noindent{\bf $($Solution in formal power series$)$.}
$$
\begin{array}{l}
u(x, t) = 4 \dst\sum^{\infty}_{n = 1} \int \frac{S(k_1, t) S(k_2, t)
\ldots S(k_n, t) (k_1 + \ldots + k_n)}{(k_1 + k_2 + i0)(k_2 + k_3 +
i0) \ldots (k_{n - 1} + k_n + i0)}
\\ [12pt]
\exp(2i(k_1 + \ldots + k_n)x)\ \dst\frac{dk_1 \ldots dk_n}{(2\pi i)^n}
\end{array}
$$
where

\vspace{6pt}
$S(k, t) = S(k) \exp(8ik^3 t)$

$$
\begin{array}{r@{\;}l}
S(k) =& -\dst\frac{1}{2i(k + i0)} \sum^{\infty}_{n = 0} \int u(x_1)
u(x_3) \ldots u(x_{2n + 1})
\\ [12pt]
& \quad \Theta_{21} \Theta_{23} \Theta_{43} \Theta_{45}
\cdot \ldots \cdot \Theta_{2n, 2n - 1} \Theta_{2n, 2n + 1}
\\ [12pt]
& \quad \delta(-x + x_1 - x_2 + x_3 - \ldots + x_{2n + 1})
\exp(-2ikx)\ dx dx_1 \ldots dx_{2n + 1}
\\ [12pt]
u(x) \equiv& u(x, 0)
\end{array}
$$
is the solution of Cauchy problem.

\noindent {\bf $($Solution in convergent series$)$.}

\noindent$1)$ Starting from $u(x)$ define $S(k)$, $\{ i \kappa_n \}^N_{n = 1}$,
$\{ c_n \}^N_{n = 1}$ as follows$:$

$S(k) = \dst\frac{b(k)}{a(k)},$ $k \in {\Bbb R},$ where
$$
\begin{array}{r@{\;}l}
a(k) =& 1 + \dst\sum^{\infty}_{n = 1} (-)^n \int u(x_2) u(x_4) \ldots
u(x_{2n}) \exp(+2ikx_0)
\\ [12pt]
& \quad \delta(x_0 - x_1 + x_2 - \ldots + x_{2n})
\Theta_{12} \Theta_{23} \ldots \Theta_{2n - 1, 2n}
\\ [12pt]
& \quad dx_0 dx_1 \ldots dx_{2n}, \quad k \in {\Bbb C}^+
\\ [12pt]
b(k) =& \dst\frac{1}{2i(k + i0)} \sum^{\infty}_{n = 0} (-)^{n + 1}
\int u(x_1) u(x_3) \ldots u(x_{2n + 1}) \exp(-2ikx_0)
\\ [12pt]
& \quad \delta(-x_0 + x_1 - x_2 + x_3 - \ldots + x_{2n + 1})
\Theta_{12} \Theta_{23} \ldots \Theta_{2n, 2n + 1}
\\ [12pt]
& \quad dx_0 dx_1 \ldots dx_{2n + 1}
\end{array}
$$
$S(k) = \bar{S}(-k); |S(k)| \leq 1, k \ne 0$
\vspace{4pt}

$\{ i \kappa_n \}^N_{n = 1}, \kappa_n \in {\Bbb R}^+$ is the set of zeroes of
the function $a(k), k \in {\Bbb C}^+$
\vspace{4pt}

$c_n = \dst\frac{\tilde{c}_n}{\frac{\partial}{\partial k} a(k) \mid_{k =
i \kappa_n}},$ where
\vspace{4pt}

$\tilde{c}_n = \dst\frac{e^{\kappa_n x} \Phi(i \kappa_n, x)}{\Psi(i
\kappa_n, x)} = \frac{\Phi(i \kappa_n, 0)}{\Psi(i \kappa_n, 0)}$ $($the
ratio doesn't depend on $x$$)$
\vspace{4pt}

$\Phi(k, x)$ and $\Psi(k, x)$ defined in $(3),$ $(4)$.
\vspace{4pt}

Define
$$
\begin{array}{r@{\;}l}
S(x, t) =& \dst\int S(k) \exp (2ikx + 8ik^3t) \frac{dk}{\pi}
\\ [12pt]
c_n(t) =& c_n \exp({8\kappa_n}^3 t)
\end{array}
$$

For the class of initial data $u(x)$ such that the function $S(x)$ is
of fast decrease as $x \rightarrow +\infty$, the solution could be
written as follows$:$

Define the Fredholm determinant and the $1$st minor to be
$$
\begin{array}{l}
D_{x, t} = 1 + \dst\sum^{\infty}_{n = 1} \frac{1}{n!}
\int^{\infty}_{-\infty} \Theta(-y_1) \Theta(-y_2) \ldots \Theta(-y_n)
\ldots \det_{{ij}} [S(x - y_i - y_j, t)]  dy_1 dy_2 \ldots dy_n
\\ [12pt]
D_{x, t} \dst{y \choose y_0} = \Theta(-y) S(x - y - y_0, t) +
\dst\sum^{\infty}_{n = 1} \frac{1}{n!} \Theta(-y) \int \Theta(-y_1)
\Theta(-y_2) \ldots \Theta(-y_n)
\\ [12pt]
\cdot
\left|
\begin{array}{llcl}
S(x - y - y_0, t)& S(x - y - y_1, t)& \ldots& S(x - y - y_n, t)
\\ [6pt]
S(x - y_1 - y_0, t)& S(x - y_1 - y_1, t)& \ldots& S(x - y_1 - y_n, t)
\\ [6pt]
&\multicolumn{1}{c}{\cdots}
\\ [6pt]
S(x - y_n - y_0, t)& S(x - y_n - y_1, t)& \ldots& S(x - y_n - y_n, t)
\end{array}
\right|  \cdot dy_1 dy_2 \ldots dy_n
\end{array}
$$
$$
\begin{array}{r@{\;}l}
u(x, t) =& -\dst\frac{\partial}{\partial x} \int S(x - y, t) \left(
\delta(y) - \frac{D_{x, t} \dst{y \choose 0}}{D_{x, t}} \right) dy
\quad -2 \dst\frac{\partial^2}{\partial x^2} \ln\det A(x, t)
\\ [12pt]
A(x, t)_{mn} =& \delta_{mn} - \dst\frac{i c_n(t) e^{-(\kappa_n +
\kappa_m)x}}{(\kappa_n + \kappa_m)} \quad +
\\ [12pt]
& 2i c_n(t) e^{-(\kappa_n +\kappa_m)x}\dst\int e^{2(\kappa_m y_0 + \kappa_n
y_1)} \frac{D_{x, t} \dst{y_0
\choose y_1}}{D_{x, t}} \Theta(-y_1) dy_0 dy_1.
\end{array}
$$
This solution is defined for the class of initial data such that the
Fredholm determinant and $1$st minor are convergent.For convergence it is
enough to have $S(x)$ of fast decrease as $x \rightarrow +\infty$. It can
be proved that the Fredholm determinant is not zero.

\vspace{6pt}
\section{Nonlocal transformations for the\hfil\break Nonlinear Schrodinger
equation\hfil\break (defocusing case)}
\setcounter{equation}{0}

Let $q(x)$ be a $C^{\infty}$ complex-valued function of a real
variable $x$, with fast decrease as $x \rightarrow \pm \infty$
(Schwarz class).

{}From $q(x)$ construct the following series:
\begin{equation}
\begin{array}{l}
a(x) = \delta(x) + \dst\sum^{\infty}_{n = 1} \dst\int^{\infty}_{-\infty}
\bar{q}(x_1) q(x_2) \bar{q}(x_3) \ldots q(x_{2n}) \Theta_{21} \Theta_{32}
\Theta_{43} \ldots \Theta_{2n, 2n - 1}
\\ [12pt]
\qquad\qquad \delta(x + x_1 - x_2 + x_3 - \ldots + x_{2n})\ dx_1 dx_2
\ldots dx_{2n}
\end{array}
\end{equation}

\begin{equation}
\begin{array}{l}
b(x) = \dst\sum^{\infty}_{n = 0} \dst\int^{\infty}_{-\infty}
q(x_1) \bar{q}(x_2) q(x_3) \ldots \bar{q}(x_{2n}) q(x_{2n + 1})
\Theta_{21} \Theta_{32} \Theta_{43} \cdot \ldots \cdot \Theta_{2n + 1, 2n}
\\ [12pt]
\qquad\qquad \delta(x - x_1 + x_2 - \ldots + x_{2n} - x_{2n + 1})
\ dx_1 dx_2 \ldots dx_{2n + 1}
\end{array}
\end{equation}

\begin{equation}
\begin{array}{l}
\Phi_1(x, y) = \delta(x) + \dst\sum^{\infty}_{n = 1}
\dst\int^{\infty}_{-\infty} q(x_1) \bar{q}(x_2) q(x_3) \ldots \bar{q}(x_{2n})
\Theta(y - x_1) \Theta_{12} \Theta_{23} \ldots \Theta_{2n + 1, 2n}
\\ [12pt]
\qquad\qquad \cdot \delta(x - x_1 + x_2 - x_3 + \ldots - x_{2n - 1} +
x_{2n})\ dx_1 \ldots dx_{2n}
\end{array}
\end{equation}

\begin{equation}
\begin{array}{l}
\Phi_2(x, y) = \dst\sum^{\infty}_{n = 0}
\dst\int^{\infty}_{-\infty} \bar{q}(x_1) q(x_2) \bar{q}(x_3) \ldots
q(x_{2n}) \bar{q}(x_{2n + 1})
\Theta(y - x_1) \Theta_{12} \Theta_{23} \ldots \Theta_{2n, 2n + 1}
\\ [12pt]
\qquad\qquad \cdot \delta(x - y + x_1 - x_2 + \ldots - x_{2n} + x_{2n
+ 1})\ dx_1 \ldots dx_{2n + 1}
\end{array}
\end{equation}

\begin{equation}
\begin{array}{l}
\Psi_1(x, y) = -\dst\sum^{\infty}_{n = 0}
\dst\int^{\infty}_{-\infty} q(x_1) \bar{q}(x_2) q(x_3) \ldots
\bar{q}(x_{2n}) q(x_{2n + 1})
\Theta(x_1 - y) \Theta_{21} \Theta_{32} \ldots \Theta_{2n + 1, 2n}
\\ [12pt]
\qquad\qquad \delta(x + y - x_1 + x_2 - x_3 + \ldots - x_{2n + 1})
\ dx_1 \ldots dx_{2n + 1}
\end{array}
\end{equation}

\begin{equation}
\begin{array}{l}
\Psi_2(x, y) = \delta (x) +
\dst\sum^{\infty}_{n = 1}\dst\int^{\infty}_{-\infty}
\bar{q}(x_1) q(x_2) \bar{q}(x_3) \ldots q(x_{2n})
\Theta(x_1 - y) \Theta_{21} \Theta_{32} \ldots \Theta_{2n, 2n - 1}
\\ [12pt]
\qquad\qquad \delta(x + x_1 - x_2 + x_3 - \ldots + x_{2n - 1} -
x_{2n})\ dx_1 \ldots dx_{2n}
\end{array}
\end{equation}

The integration domain, say for $n$th term in $\Psi_1(x, y)$, is the
intersection of the region $y \leq x_1 \leq x_2 \leq \ldots \leq
x_{2n + 1}$ with the hyperplane $x + y - x_1 + x_2 - x_3 + \ldots -
x_{2n + 1} = 0$.

{\bf Lemma}.
\begin{em}
The series (1)--(6) are convergent.
\end{em}

\vspace{6pt}

Consider, for example, the series for $\Phi_1(x, y)$. Let $Q =
\max|q(x)|$.
$$
\begin{array}{l}
\bigl| \dst\int q(x_1) \bar{q}(x_2) q(x_3) \ldots \bar{q}(x_{2n})
\Theta(y - x_1) \Theta_{12} \Theta_{23} \cdot \ldots \cdot \Theta_{2n
- 1, 2n} \bigr.
\\ [12pt]
\quad \delta(x - x_1 + x_2 - \ldots + x_{2n})\ dx_1 \ldots dx_{2n}
\\ [12pt]
\leq Q \cdot \dst\int |q(x_1)| |q(x_2)| \ldots |q(x_{2n - 1})| \Theta_{12}
\Theta_{23} \ldots \Theta_{2n - 1, 2n}\ dx_1 \ldots dx_{2n - 1}
\\ [12pt]
= \dst\frac{Q}{(2n - 1)!} \left( \int^{\infty}_{-\infty} |q(x)| dx
\right)^{2n - 1}.
\end{array}
$$
Define $S(x) = \dst\int \frac{b(k)}{a(k)} e^{2ikx} \frac{dk}{\pi}$,
where
\begin{equation}
b(k) = \int b(x) e^{-2ikx}\ dx, \quad a(k) = \int a(x) e^{-2ikx}\ dx
\end{equation}

\vspace{4pt}

{\bf Lemma}.

\noindent 1).For $q(x)$ such that $|1 - a(k)| < 1$ $S(x)$ is given by the
convergent
series
\begin{equation}
\begin{array}{l}
S(x) = \dst\sum^{\infty}_{n = 0} (-)^n \dst\int
q(x_1) \bar{q}(x_2) q(x_3) \ldots \bar{q}(x_{2n}) q(x_{2n + 1})
\\ [12pt]
\qquad\qquad \Theta_{12} \Theta_{32} \Theta_{34} \Theta_{54} \cdot \ldots \cdot
\Theta_{2n - 1, 2n} \Theta_{2n + 1, 2n}
\\ [12pt]
\qquad\qquad \cdot \delta(x - x_1 + x_2 - \ldots + x_{2n} - x_{2n + 1})
\ dx_1 dx_2 \ldots dx_{2n + 1}
\end{array}
\end{equation}

\noindent 2).Consider $S(x)$ ,given by the formal series (8). We can prove the
following relation:

$$
\int S(x_1) a(x - x_1) dx_1 = b(x).
$$

\noindent Here $a(x)$ and $b(x)$ are functionals in $q(x), \bar{q}(x)$, given
by the series (1) and (2).

Indeed$,$ let us collect all
the terms of the same order in $q$, $\bar{q}$ in the convolution of
the series $(1)$ and $(8)$
$$
\begin{array}{l}
\dst\int S(y) a(x - y) dy = \dst\int q(x_1) \delta(y - x_1) \delta(x
- y) dx_1 dy + \int q(x_1) \bar{q}(x_2) q(x_3)
\\ [12pt]
\quad \left( -\Theta_{12} \Theta_{32} \delta(y - x_1 + x_2 - x_3)
 \delta(x - y) + \right.
\\ [12pt]
\quad  \Theta_{32} \delta(y - x_1) \delta(x - y +
 x_2 - x_3)\ dx_1 dx_2 dx_3 dy
\end{array}
$$
$$
\begin{array}{l}
+ \ldots + \dst\int q(x_1) \bar{q}(x_2) \ldots q(x_{2n + 1}) (-)^n
\\ [12pt]
\quad \left( \Theta_{12} \Theta_{32} \Theta_{34} \Theta_{54} \ldots
\Theta_{2n - 1, 2n - 2} \Theta_{2n - 1, 2n} \Theta_{2n + 1, 2n}
\right.
\\ [12pt]
+ \dst\sum^{n - 1}_{m = 0} (-)^{m - n} \Theta_{12} \Theta_{32} \Theta_{34}
\Theta_{54} \ldots \Theta_{2m - 1, 2m} \Theta_{2m + 1, 2m}
\\ [12pt]
\quad (\Theta_{2m + 2, 2m + 1} + \Theta_{2m + 1, 2m + 2}) \Theta_{2m + 3,
2m + 2}
\\ [12pt]
\quad \cdot \left. \Theta_{2m + 4, 2m + 3} \cdot \ldots \cdot \Theta_{2n +
1, 2n} \right) \delta(x - x_1 + x_2 - x_3 + \ldots - x_{2n + 1})\ dx_1
\ldots dx_{2n}
\\ [12pt]
= \dst\sum^{\infty}_{n = 0} \int q(x_1) \bar{q}(x_2) \ldots q(x_{2n +
1}) \Theta_{21} \Theta_{32} \Theta_{43} \cdot \ldots \cdot \Theta_{2n
+ 1, 2n}
\\ [12pt]
\quad \delta(x - x_1 + x_2 - x_3 + \ldots - x_{2n + 1})\ dx_1 \ldots
dx_{2n}
\\ [12pt]
= b(x)
\end{array}
$$
$($We used $\Theta_{ij} + \Theta_{ji} = 1$$)$.
\vskip18pt

The convolutions of functionals $(1)$--$(8)$ are again the
functionals of the same type. There are certain relations for the
convolutions:

\begin{equation}
\dst\int \bar{a}(x_1) a(x + x_1) dx_1 = \delta(x) + \dst\int
\bar{b}(x_1) b(x + x_1)\ dx_1
\end{equation}

\begin{equation}
\dst\int \left( b(x_1) \Phi_2 (x - x_1 + y, y) - a(x_1)
\bar{\Phi}_1 (-x + x_1, y) \right) dx_1 = -\Psi_2(x, y)
\end{equation}

\begin{equation}
\dst\int \left( b(x_1) \Phi_1 (x - x_1 + y, y) - a(x_1)
\bar{\Phi}_2 (-x + x_1, y) \right) dx_1 = -\Psi_1(x, y)
\end{equation}
\\ [12pt]

Let us prove (10). We have to collect all the terms of the same degree
in $q$, $\bar{q}$ in the left-hand side and to compare with the right-hand
side.
$$
\begin{array}{l}
\dst\int (\Phi_2 (x - s + y, y) b(s) - \bar{\Phi}_1 (-x + s,
y) a(s))\ ds
\\ [12pt]
= -\delta(x) + \dst\int \bar{q}(x_1) q(x_2) \left( \Theta(y - x_1) \delta(x
- s + x_1) \delta(s - x_2) - \Theta(y - x_1) \Theta_{12} \right.
\\ [12pt]
\left. \delta(x - s + x_1 - x_2) d(s) \cdot -\Theta_{21} \delta(x - s)
\delta(s + x_1 - x_2) \right) dx_1 dx_2 ds + \ldots
\\ [12pt]
+ \dst\int \bar{q}(x_1) q(x_2) \bar{q}(x_3) \ldots q(x_{2n})
\\ [12pt]
\cdot \left( \dst\sum^{n - 1}_{m = 0} \Theta(y - x_1) \Theta_{12}
\Theta_{23} \Theta_{34} \cdot \ldots \cdot \Theta_{2m, 2m + 1}
(\Theta_{2m + 1, 2m + 2} + \Theta_{2m + 2, 2m + 1}) \right.
\end{array}
$$
$$
\begin{array}{l}
\Theta_{2m + 3, 2m + 2} \Theta_{2m + 4, 2m + 3} \cdot \ldots \cdot
\Theta_{2n, 2n - 1} - \Theta_{21} \Theta_{32} \cdot \ldots \cdot
\Theta_{2n, 2n - 1}
\\ [12pt]
-\dst\sum_{m = 1} \Theta(y - x_1) \Theta_{12} \Theta_{23} \cdot
\ldots \cdot \Theta_{2m - 1, 2m} (\Theta_{2m, 2m + 1} + \Theta_{2m +
1, 2m})
\\ [12pt]
\Theta_{2m + 2, 2m + 1} \cdot \Theta_{2m + 3, 2m + 2} \cdot \ldots
\cdot \Theta_{2n, 2n - 1}
\\ [12pt]
- \left. \Theta(y - x_1) \Theta_{12} \Theta_{23} \cdot \ldots \cdot
\Theta_{2n, 2n - 1} \right) \delta(x + x_1 - x_2 + \ldots - x_{2n})
\ dx_1 \ldots dx_{2n}
\\ [12pt]
= -\delta(x) - \dst\sum^{\infty}_{n = 1} \int \bar{q}(x_1) q(x_2)
\bar{q}(x_3) \ldots q(x_{2n}) \Theta(x_1 - y) \Theta_{21} \Theta_{32}
\cdot \ldots \cdot \Theta_{2n, 2n - 1}
\\ [12pt]
\delta(x + x_1 - x_2 + \ldots - x_{2n}) dx_1 \ldots dx_{2n} = -\Psi_2
(x, y)
\end{array}
$$

\vspace*{-2pt}
The proof of the other relations is similar.

In addition to convolution of two functionals, there is another operation for
our functionals , namely inversion.It is the
infinite-dimensional analoque of the inverse function. Consider the series (8)
\vspace*{-6pt}
$$
\begin{array}{r@{\;}l}
S(x) =& \dst\sum^{\infty}_{i = 0} q_{(2i + 1)} (x) :=
\\ [9pt]
&\quad = q(x) - \dst\int q(x_1) \bar{q}(x_2) q(x_3) \Theta_{12}
\Theta_{32} \delta(x - x_1 + x_2 - x_3)
\\ [9pt]
&\quad \phantom{= q(x)} + \dst\int q(x_1) \bar{q}(x_2) q(x_3)
\bar{q}(x_4) q(x_5) \Theta_{12} \Theta_{32} \Theta_{34} \Theta_{54}
\\ [9pt]
&\quad \phantom{= q(x)}\ \delta(x - x_1 + x_2 - x_3 + x_4 - x_5) - \ldots
\end{array}
$$
\vspace*{-6pt}

$S(x)$ is a formal series , its $n$th term to is a nonlocal analytic
functional of $q(x)$ , $\bar{q}(x)$ of degree $(2n + 1)$ in $q,\bar{q}$. It can
be inverted,
namely, $q(x)$ can be expressed in terms of $S(x)$:
$$
q(x) = S_{(1)}(x) + S_{(3)}(x) + S_{(5)}(x) +\ldots
$$

where $S_{m}(x)$ is a nonlocal analytic functional of $S(x),\bar{S}(x)$ of
degree m in $S,\bar{S}$.
$$
\begin{array}{l}
S_{(1)}(x) = S(x)
\\ [9pt]
S_{(3)}(x) = \dst\int S(x_1) \bar{S}(x_2) S(x_3) \delta(x - x_1 + x_2
- x_3) \Theta_{12} \Theta_{32}\ dx_1 dx_2 dx_3
\\ [9pt]
S_{(5)}(x) = -\dst\int S(x_1) \bar{S}(x_2) S(x_3) \bar{S}(x_4) S(x_5)
\Theta_{12} \Theta_{32} \Theta_{34} \Theta_{54}
\\ [9pt]
\qquad\qquad \delta(x - x_1 + x_2 - x_3 + x_4 - x_5)\ dx_1 \ldots dx_5
\\ [9pt]
\qquad\qquad + \dst\int \left( S^{(3)}_{(3)} (x_1) \bar{S}(x_2) S(x_3) + S(x_1)
\right.
\\ [9pt]
\qquad\qquad \left. \bar{S}^{(3)}_{(3)}(x_2) S(x_3) + S(x_1)
\bar{S}(x_2) S^{(3)}_{(3)} (x_3) \right)
\\ [9pt]
\qquad\qquad \Theta_{12} \Theta_{32} \delta(x - x_1 + x_2 - x_3)\
dx_1 dx_2 dx_3
\\ [9pt]
\phantom{S_{(5)}(x)} = \dst\int S(x_1) \bar{S}(x_2) S(x_3)
\bar{S}(x_4) S(x_5) \delta(x - x_1 + x_2 - x_3 + x_4 - x_5)
\end{array}
$$
$$
\begin{array}{l}
\left(\ - \Theta_{12} \Theta_{32} \Theta_{34} \Theta_{54} \right.
\\ [9pt]
\phantom{(} + \Theta_{12} \Theta_{32} \phantom{\Theta_{34}} \Theta_{54}
\Theta(x_1 - x_2 + x_3 - x_4)
\\ [9pt]
\phantom{\,\, (+ \Theta_{12}} \Theta_{23} \Theta_{43}
\phantom{\Theta_{54}} \Theta(x_1 - x_2 + x_3 - x_4) \Theta(x_5 - x_4
+ x_3 - x_2)
\\ [9pt]
\left. \phantom{(} + \Theta_{12} \phantom{\Theta_{23}} \Theta_{34} \Theta_{54}
\Theta(x_5 - x_4 + x_3 - x_2) \right) dx_1 \ldots dx_5
\\ [9pt]
= \dst\int S(x_1) \bar{S}(x_1) S(x_3) \bar{S}(x_4) S(x_5) \delta(x -
x_1 + x_2 - x_3 + x_4 - x_5)
\\ [9pt]
\cdot \Theta_{12} \Theta(x_1 - x_2 + x_3 - x_4) \Theta_{54}
\Theta(x_5 - x_4 + x_3 - x_2)
\\ [9pt]
(-\Theta_{32} \Theta_{34} + \Theta_{32} +
\Theta_{23} \Theta_{43} + \Theta_{34})\ dx_1 \ldots dx_5
\\ [9pt]
= \dst\int S(x_1) \bar{S}(x_2) S(x_3) \bar{S}(x_4) S(x_5) \Theta_{12}
\Theta(x_1 - x_2 + x_3 - x_4) \Theta_{54}
\\ [9pt]
\cdot \Theta(x_5 - x_4 + x_3 - x_2) \delta(x - x_1 + \ldots - x_5)
\ dx_1 \ldots dx_5
\end{array}
$$
(Indeed, $-\Theta_{32} \Theta_{34} + \Theta_{32} + \Theta_{23} \Theta_{43} +
\Theta_{34} = -\Theta_{32} \Theta_{34} + \Theta_{32} (\Theta_{34} +
\Theta_{43}) + \Theta_{23} \Theta_{43} + \Theta_{34} = (\Theta_{23} +
\Theta_{32}) \Theta_{43} + \Theta_{34} = \Theta_{43} + \Theta_{34} = 1$.

\vspace{4pt}

{\bf Lemma}.
\begin{em}
Consider the nonlocal analytic functionals of $\{ S(x) \},$ given by
formal series
\begin{equation}
\begin{array}{l}
\tilde{q}(x) = S(x) + \dst\sum^{\infty}_{n = 1} \int \left( S(x_1)
\bar{S}(x_2) S(x_3) \ldots \bar{S}(x_{2n}) S(x_{2n + 1}) \right.
\\ [9pt]
\qquad \Theta(x_1 - x_2) \Theta(x_1 - x_2 + x_3 - x_4) \ldots
\Theta(x_1 - x_2 + \ldots - x_{2n})
\\ [9pt]
\qquad \Theta(x_{2n + 1} - x_{2n}) \Theta(x_{2n + 1} - x_{2n} + x_{2n
- 1} - x_{2n - 2}) \ldots \Theta(x_{2n + 1} - x_{2n} + \ldots - x_2)
\\ [9pt]
\qquad \delta(x - x_1 + x_2 - \ldots + x_{2n} - x_{2n + 1})\ dx_1 \ldots
dx_{2n + 1}
\end{array}
\end{equation}
\begin{equation}
\begin{array}{l}
\tilde{\Phi}_1(x, y) = \delta(x) + \dst\sum^{\infty}_{n = 1} \int
\left( S(x_1) \bar{S}(x_2) S(x_3) \ldots \bar{S}(x_{2n}) \right.
\\ [9pt]
\qquad \cdot \Theta(y - x_1) \Theta(y - x_1 + x_2 - x_3) \ldots
\Theta(y - x_1 + \ldots - x_{2n - 1})
\\ [9pt]
\qquad \cdot \Theta(x_1 - x_2) \Theta(x_1 - x_2 + x_3 - x_4) \cdot
\ldots \cdot \Theta(x_1 - x_2 + \ldots - x_{2n})
\\ [9pt]
\qquad \delta(x - x_1 + x_2 - x_3 + \ldots - x_{2n + 1} + x_{2n})\ dx_1 \ldots
dx_{2n}
\end{array}
\end{equation}
\begin{equation}
\begin{array}{l}
\tilde{\Phi}_2(x, y) = \dst\sum^{\infty}_{n = 0} \int
\bar{S}(x_1) S(x_2) \bar{S}(x_3) \ldots S(x_{2n}) \bar{S}(x_{2n + 1})
\\ [9pt]
\qquad \cdot \Theta(y - x_1) \cdot \Theta(y - x_1 + x_2 - x_3) \cdot \ldots
\cdot \Theta(y - x_1 + x_2 - x_3 + \ldots - x_{2n - 1})
\\ [9pt]
\qquad \cdot \Theta(x_1 - x_2) \Theta(x_1 - x_2 + x_3 - x_4) \cdot
\ldots \cdot \Theta(x_1 - x_2 + \ldots - x_{2n})
\\ [9pt]
\qquad \delta(x - y + x_1 - x_2 + \ldots - x_{2n} + x_{2n + 1})\ dx_1 \ldots
dx_{2n + 1}
\end{array}
\end{equation}

Let us substitute in these series $S(x)$ as formal series in $\{
q(x) \}$, $(8)$. The result of such substitution would be formal
series in $\{ q(x) \},$ and $,$ moreover$,$
$$
\begin{array}{l}
\tilde{q}(x) = q(x)
\\ [9pt]
\tilde{\Phi}_1(x, y) = \Phi_1(x, y)
\\ [9pt]
\tilde{\Phi}_2(x, y) = \Phi_2(x, y)
\end{array}
$$

\noindent as formal series in ${q(x)}$ , with $\Phi_1(x,y)$ and $\Phi_2(x,y)$
given by (3) , (4).
\end{em}

The series (9) could be used to get the solution of a nonlinear equation.
In order to see this, let us rewrite (9) in terms of
$ S(k)=\dst\int^{\infty}_{-\infty}S(x)e^{-2ikx}dx$ and
$\bar{S}=(k)\dst\int^{\infty}_{-\infty}\bar{S}(x)e^{2ikx}dx$ :

\begin{equation}
\begin{array}{l}
q(x)=2\dst\sum^{\infty}_{n = 0}\dst\int^{\infty}_{-\infty}
\frac{S(k_1)
\bar{S}(k_2)\ldots S(k_{2n+1})}{(k_2 - k_1 + i0)(k_3 - k_2 - i0)(k_4 - k_3 +
i0) \ldots
(k_{2n+1} - k_{2n} - i0)}
\\ [12pt]
\qquad \cdot \exp(2i(k_1 - k_2 + k_3 - \ldots + k_{2n+1})x)\ \dst\frac{dk_1
\ldots
dk_{2n+1}}{(2\pi)^{2n+1}}
\end{array}
\end{equation}

The kernels
$\dst\frac{1}{k \mp i0}=2i\dst\int^{\infty}_{-\infty}\Theta(x)exp(\mp2ikx)$
appeared as the Fourier transform of the Heaviside function kernels.

The series (15) has the following property: a polynomial in $q(x)$ ,
$\bar{q}(x)$
and their derivatives can also be written in the form (15) but with some
polynomial in $\{k\}$ in the numerator:

\begin{equation}
\begin{array}{l}
\dst\frac{d^nq(x)}{dx^n} = 2\cdot(2i)^n \dst\sum^{\infty}_{n = 0}
\\ [12pt]
\qquad \dst\int \frac{S(k_1)
\bar{S}(k_2)  \ldots {S}(k_{2n+1}) {(k_1 - k_2 + \ldots -
k_{2n} + k_{2n+1})}^{n}}{(k_2 - k_1 + i0)(k_3 - k_2 - i0)(k_4 - k_3 + i0)
\ldots
(k_{2n+1} - k_{2n} - i0)}
\\ [12pt]
\qquad \cdot \exp(2i(k_1 - k_2 + k_3 - \ldots + k_{2n+1})x)\ \dst\frac{dk_1
\ldots
dk_{2n+1}}{(2\pi)^{2n+1}}
\\ [30pt]

q(x) \bar{q}(x) = 4 \dst\sum^{\infty}_{n = 1}
\\ [12pt]
\qquad \dst\int \frac{S(k_1)
\bar{S}(k_2) S(k_3) \ldots \bar{S}(k_{2n}) (-k_1 + k_2 - \ldots +
k_{2n})}{(k_2 - k_1 + i0)(k_3 - k_2 - i0)(k_4 - k_3 + i0) \ldots
(k_{2n - 1} - k_{2n - 2} - i0)(k_{2n} - k_{2n - 1} + i0)}
\\ [12pt]
\qquad \cdot \exp(2i(k_1 - k_2 + \ldots - k_{2n})x)\ \dst\frac{dk_1 \ldots
dk_{2n}}{(2\pi)^{2n}}
\\ [30pt]

q(x) \bar{q}(x) q(x) = 4 \dst\sum^{\infty}_{n = 1}
\\ [12pt]
\qquad \dst\int \frac{S(k_1)
\bar{S}(k_2) S(k_3) \ldots \bar{S}(k_{2n}) S(k_{2n + 1})}{(k_2 - k_1
+ i0)(k_3 - k_2 - i0) \ldots (k_{2n} - k_{2n - 1} - i0)(k_{2n + 1} -
k_{2n} + i0)}
\\ [12pt]
\qquad \cdot \left( -(k_1 - k_2 + \ldots - k_{2n + 1})^2 - \dst\sum^n_{p =
1} {k_{2p}}^2 + \sum^n_{p = 0} {k_{2p + 1}}^2 \right)
\\ [12pt]
\qquad \exp(2i(k_1 - k_2 + \ldots - k_{2n} + k_{2n + 1}))
\ \dst\frac{dk_1 \ldots dk_{2n + 1}}{(2\pi)^{2n + 1}}
\\ [30pt]
q^{m + 1}(x) {\bar{q}}^{m}(x) = 2^{2m + 1} \dst\sum^{\infty}_{n = m}
\\ [12pt]
\qquad \left( \dst\int \frac{S(k_1) \bar{S}(k_2) S(k_3) \ldots
\bar{S}(k_{2n}) S(k_{2n + 1})}{(k_2 - k_1 - i0)(k_3 - k_2 + i0) \ldots
(k_{2n} - k_{2n - 1} - i0)(k_{2n + 1} - k_{2n} + i0)} \right.
\\ [12pt]
\qquad \cdot \dst\sum_{1 \leq p_1 < p_2 < \ldots < p_m \leq n} (-k_1 + k_2 -
\ldots + k_{2p_1})
\\ [12pt]
\qquad (k_{2p_1 + 1} - k_{2p_1})(-k_{2p_1 + 1} + k_{2p_1 +
2} - \ldots + k_{2p_2})(k_{2p_2 + 1} - k_{2p_2}) \cdot \ldots
\\ [12pt]
\qquad \cdot (-k_{2p_m + 1} + k_{2p_m + 2} - \ldots + k_{2p_m})(k_{2p_m + 1}
- k_{2p_m})\ \dst\frac{dk_1 \ldots dk_{2n + 1}}{(2\pi)^{2n + 1}}
\end{array}
\end{equation}

The fact that both differentiation of $q(x)$ and nonlinearity under
the ``nonlinear Fourier transformation'' $(15)$ has the same
effect$,$ namely$,$ some polynomial in $\{ k \}$ appears in the
numerator$,$ can be used to solve nonlinear equations.

{\bf Theorem 3.1 (solution of the NLS Equation).}

Consider the Cauchy problem for the NLS equation:
\begin{equation}
\begin{array}{l}
i \dst\frac{\partial}{\partial t} q(x, t) + \frac{\partial^2}{\partial
x^2} q(x, t) - 2\lambda |q|^2 q = 0, \quad \lambda = \left\{
\begin{array}{rl} 1,& \mbox{defocusing case}\\ -1,& \mbox{focusing
case} \end{array} \right.
\\ [12pt]
q(x, 0) = q(x)
\end{array}
\end{equation}

{\bf (Solution in formal series).}

The solution of Cauchy problem is given by
\begin{equation}
\begin{array}{r@{\;}l}
q(x, t) =& 2 \dst\sum^{\infty}_{n = 0} \lambda^n \int \frac{S(k_1, t)
\bar{S}(k_2, t) \ldots S(k_{2n + 1},t)}{(k_2 - k_1 + i0)(k_3 - k_2 -
i0) \ldots (k_4 - k_3 + i0)(k_{2n + 1} - k_{2n} - i0)}
\\ [12pt]
& \qquad\qquad \exp (2i(k_1 - k_2 + \ldots - k_{2n} + k_{2n + 1})
\\ [12pt]
& \qquad\qquad \dst\frac{dk_1 \ldots dk_{2n + 1}}{(2\pi)^{2n + 1}}
\end{array}
\end{equation}

\begin{equation}
\begin{array}{r@{\;}l}
S(k, t) =& e^{-4ik^2t} S(k)
\\ [12pt]
S(k) =& \dst\sum^{\infty}_{n = 0} (-)^n \lambda^n
\int^{\infty}_{-\infty} q(x_1) \bar{q}(x_2) q(x_3) \ldots
\bar{q}(x_{2n}) q(x_{2n + 1})
\\ [12pt]
& \qquad \Theta_{12} \Theta_{32} \Theta_{34} \Theta_{54} \ldots
\Theta_{2n - 1, 2n} \Theta_{2n + 1, 2n}
\\ [12pt]
& \qquad \exp(-2ikx) \delta(x - x_1 + x_2 - \ldots + x_{2n} - x_{2n +
1})\ dx dx_1 dx_2 \ldots dx_{2n + 1}
\end{array}
\end{equation}

\vspace{4pt}

{\bf (Solution in convergent series, defocusing case $\lambda = 1$).}

$$
\begin{array}{l}
q(x, t) = \dst\int S(x - y, t) \left( \delta(y) - \dst\frac{D_{x,
t}\dst{y \choose 0}}{D_{x, t}} \right) dy
\end{array}
$$
$S(x, t) = \dst\int \frac{b(k)}{a(k)} \exp(2ikx - 4ik^2 t)
\frac{dk}{\pi},$ $b(k)$ and $a(k)$ are defined by $(1)$ $(2)$
$$
\begin{array}{l}
D_{x, t} \dst{y \choose y_0} = -\Theta(-y) K_{x, t} (y, y_0) +
\dst\sum^{\infty}_{n = 1} \frac{(-)^{n + 1}}{n!}
\\ [12pt]
\Theta(-y) \dst\int \Theta(-y_1) \Theta(-y_2) \ldots \Theta(-y_n)
\\ [12pt]
\cdot \left|
\begin{array}{llcl}
K_{x, t} (y, y_0)& K_{x, t} (y, y_1)& \ldots& K_{x, t} (y, y_n)
\\ [6pt]
K_{x, t} (y_1, y_0)& K_{x, t} (y_1, y_1)& \ldots& K_{x, t} (y_1, y_n)
\\ [6pt]
& \multicolumn{1}{c}{\cdots}
\\ [6pt]
K_{x, t} (y_n, y_0)& K_{x, t} (y_n, y_1)& \ldots& K_{x, t} (y_n, y_n)
\end{array}
\right| \ dy_1 \ldots dy_n
\\ [12pt]
D_{x, t} = 1 + \dst\sum^{\infty}_{n = 1} \frac{(-)^n}{n!} \int
\Theta(-y_1) \Theta(-y_2) \ldots \Theta(-y_n)\ |K_{x, t} (y_i,
y_j)|_{i, j}\ dy_1 \ldots dy_n
\\ [12pt]
K_{x, t} (y_1, y_2) = \dst\int \Theta(-y) \bar{S}(x - y_1 - y) S(x -
y - y_2, t)\ dy
\end{array}
$$

The solution is defined for the class of initial data $q(x)$ such
that the Fredholm determinant $D_{x, t}$ and the first minor $D_{x, t} {y
\choose
y_0}$  are convergent. It can be proved that the determinant is not
zero.

\vspace{12pt}

{\bf Proof (formal series).}

1) We compute $\dst\frac{\partial^2}{\partial x^2} q(x, t)$ and
$|q|^2 q$ in the same way as before, see (16) ;
$$
\begin{array}{l}
i \dst\frac{\partial}{\partial t} q(x, t) +
\frac{\partial^2}{\partial x^2} q(x, t) - 2|q|^2 q
\\ [12pt]
= 2 \dst\int (i \frac{\partial}{\partial t} - 4{k_1}^2) S(k, t)
e^{2ik_1 x}\ \frac{dk_1}{2\pi}
\\ [12pt]
+ \dst\sum^{\infty}_{n = 1} 2 \lambda^n (i \frac{\partial}{\partial
t} - 4({k_1}^2 - {k_2}^2 + {k_3}^2 - \ldots + {k_{2n + 1}}^2))
\\ [12pt]
\dst\frac{S(k_1, t) \bar{S}(k_2, t) \ldots S(k_{2n + 1}, t)}{(k_2 -
k_1 + i0)(k_3 - k_2 - i0) \ldots (k_{2n + 1} - k_{2n} - i0)}
\\ [12pt]
\cdot \exp(2i(k_1 - k_2 + k_3 - \ldots + k_{2n + 1})x)\ \dst\frac{dk_1
\ldots dk_{2n + 1}}{(2\pi)^{2n + 1}} = 0
\end{array}
$$

for $S(k, t) = e^{-4ik^2 t} S(k,0)$ .

\vspace{12pt}

2) The substitution of $S(k, 0) = S(k)$ as series in $\{ q(x) \}$
(see (19)) into (18 ) gives $q(x, 0) = q(x)$ (see Lemma) .

\section{Nonlinear transformations for Davey-Stewardson equation}
\setcounter{equation}{0}
Let $q(x, y)$ be a complex-valued $C^{\infty}$ function on the plane
${\Bbb R}^2$, with fast decrease as $|x|^2 + |y|^2 \rightarrow
\infty$.

We construct the following nonlocal analytic functionals of $\{ q
\}$:\linebreak[4]

\vspace*{12pt}

\begin{equation}
\begin{array}{l}
\alpha(k, \bar{k}) = \dst\sum^{\infty}_{n = 0}
\\ [12pt]
\qquad \dst\int \frac{q(z_1, \bar{z}_1) \bar{q}(z_2, \bar{z}_2)
q(z_3, \bar{z}_3) \ldots \bar{q}(z_{2n}, \bar{z}_{2n}) q(z_{2n + 1},
\bar{z}_{2n + 1})}{(\bar{z}_1 - \bar{z}_2)(z_2 - z_3)(\bar{z}_3 -
\bar{z}_4) \ldots (z_{2k} - z_{2k + 1})}
\\ [12pt]
\qquad \cdot \exp \left( \bar{k}(\bar{z}_1 - \bar{z}_2 + \bar{z}_3 -
\ldots + \bar{z}_{2k + 1}) - k(z_1 - z_2 + z_3 - \ldots + z_{2k + 1})
\right)
\\ [12pt]
\qquad \dst\frac{d^2z_1 d^2z_2 \ldots d^2z_{2n + 1}}{(2\pi)^{2n + 1}}
\end{array}
\end{equation}
\begin{equation}
\begin{array}{l}
\mu_1(k, \bar{k}; z, \bar{z}) = 1 + \dst\sum^{\infty}_{n = 1}
\\ [12pt]
\qquad \dst\int \frac{q(z_1, \bar{z}_1) \bar{q}(z_2, \bar{z}_2)
\ldots \bar{q}(z_{2n - 1}, \bar{z}_{2n - 1}) \bar{q}(z_{2n},
\bar{z}_{2n})}{(z - z_1)(\bar{z}_1 - \bar{z}_2)(z_2 - z_3) \ldots
(\bar{z}_{2n - 1} - \bar{z}_{2n})}
\\ [12pt]
\qquad \cdot \exp \left( \bar{k}(\bar{z}_1 - \bar{z}_2 + \bar{z}_3 -
\ldots + \bar{z}_{2n + 1} - \bar{z}_{2n}) - k(z_1 - z_2 + z_3 -
\ldots + z_{2n + 1} - z_{2n}) \right)
\\ [12pt]
\qquad \dst\frac{d^2z_1 \ldots d^2z_{2n}}{(2\pi)^{2n}}
\end{array}
\end{equation}
\begin{equation}
\begin{array}{l}
\mu_2(k, \bar{k}; z, \bar{z}) = \dst\sum^{\infty}_{n = 0}
\\ [12pt]
\qquad \dst\int \frac{\bar{q}(z_1, \bar{z}_1)
q(z_2, \bar{z}_2) \ldots \bar{q}(z_{2n + 1}, \bar{z}_{2n + 1})}
{(\bar{z} - \bar{z}_1)(z_1 - z_2) (\bar{z}_2 - \bar{z}_3) \ldots
(z_{2n - 1} - z_{2n})(\bar{z}_{2n} - \bar{z}_{2n + 1})}
\\ [12pt]
\qquad \cdot \exp \left( \bar{k}(\bar{z} - \bar{z}_1 + \bar{z}_2 -
\ldots + \bar{z}_{2n} - \bar{z}_{2n + 1}) - k(z - z_1 + z_2 -
\ldots + z_{2n} - z_{2n + 1}) \right)
\\ [12pt]
\qquad \dst\frac{d^2z_1 \ldots d^2z_{2n}}{(2\pi)^{2n + 1}}
\end{array}
\end{equation}
where $z = x_1 + ix_2$, $\bar{z} = x_1 - ix_2$, $k = k_1 + ik_2$,
$\bar{k} = k_1 - ik_2$, $d^2 z := \dst\frac{i}{2} dz d\bar{z}$.

The series are convergent for some class of functions $q$. We will
not investigate convergence; we will work with these functionals
as with formal series. Each term of these series is an integral,
involving homogeneous generalized functions
$\dst\frac{1}{z}$ as kernels ([1 ]).

There are the following relations for the functionals (1 )--(3 ):
$$
\begin{array}{l}
\dst\frac{\partial \mu_1}{\partial \bar{z}} (k, \bar{k}, z, \bar{z})
= \dst\frac{1}{2} q(z ,\bar{z}) \mu_2(k, \bar{k}, z, \bar{z}).
\\ [12pt]
\dst\frac{\partial \mu_2}{\partial \bar{z}} (k, \bar{k}, z, \bar{z})
= k\mu_2(k, \bar{k}, z, \bar{z}) + \dst\frac{1}{2} q(z ,\bar{z})
\mu_1(k, \bar{k}, z, \bar{z}).
\\ [12pt]
\dst\frac{\partial \mu_1}{\partial \bar{k}} = e^{\bar{k}\bar{z} - kz}
\bar{\alpha} \bar{\mu}_2 .
\\ [12pt]
\dst\frac{\partial \mu_2}{\partial \bar{k}} = e^{\bar{k}\bar{z} - kz}
\bar{\alpha} \bar{\mu}_1 .
\end{array}
$$

\vspace{2pt}
The series (1 ) could be inverted, that is, $q(z, \bar{z})$ could be
written as a functional of $\{ \alpha(k, \bar{k}) \}$:
$$
\begin{array}{l}
\alpha(k, \bar{k}) = \dst\int q(z_1, \bar{z}_1) \exp(\bar{k}\bar{z}_1 -
kz_1) \frac{d^2z_1}{2\pi} + \dst\int \left( \frac{q(z_1, \bar{z}_1)
\bar{q}(z_2, \bar{z}_2) q(z_3, \bar{z}_3)}{(z_1 - z_2)(\bar{z}_2 -
\bar{z}_3)}\right.
\\ [12pt]
\left. \exp(\bar{k}(\bar{z}_1 - \bar{z}_2 + \bar{z}_3) - k(z_1 - z_2
+ z_3)) \phantom{\dst\frac{1}{2}}\right) \dst\frac{d^2z_1 d^2z_2
d^2z_3}{(2\pi)^3} + \ldots
\\ [12pt]
q(z, \bar{z}) = \alpha_{(1)} (z, \bar{z}) + \alpha_{(3)} (z, \bar{z}) + \ldots
\\ [12pt]
\alpha_{(1)} (z, \bar{z}) = -2 \dst\int \alpha(k_1, \bar{k}_1) \exp(-\bar{k},
\bar{z} + k\bar{z})\ \frac{d^2 k}{\pi}
\end{array}
$$

\vspace{6pt}

$$
\begin{array}{r@{\;}l}
\alpha_{(3)} (z, \bar{z}) =& -\dst\frac{2}{\pi^7} \int
\frac{\alpha(k_1, \bar{k}_1) \bar{\alpha}(k_2, \bar{k}_2) \alpha(k_3,
\bar{k}_3)}{(z_1 - z_2)(\bar{z}_2 - \bar{z}_3)}
\\ [12pt]
& \quad \exp \left((-\bar{k}_1 \bar{z}_1 + \bar{k}_2 \bar{z}_2 - \bar{k}_3
z_3 + \bar{k}(\bar{z}_1 - \bar{z}_2 + \bar{z}_3) - \bar{k} \bar{z}) \right.
\\ [12pt]
& \quad d^2z_1 d^2z_2 d^2z_3 d^2k_1 d^2k_2 d^2k_3 d^2k
\\ [12pt]
=& -\dst\frac{2}{\pi^3} \int \frac{\alpha(k_1, \bar{k}_1)
\bar{\alpha}(k_2, \bar{k}_2) \bar{\alpha}(k_3, \bar{k}_3)}{(\bar{k} -
\bar{k}_1)(k_2 - k_1)}\ \delta(k - k_1 + k_2 - k_3)
\\ [12pt]
& \quad \exp(-\bar{k} \bar{z} + kz) d^2k_1 d^2k_2 d^2k_3 d^2k_4
\\ [12pt]
=& -2 \dst\int \frac{\alpha(k_1, \bar{k}_1) \bar{\alpha}(k_2,
\bar{k}_2) \alpha(k_3, \bar{k}_3)}{(k_2 - k_1)(\bar{k}_3 - \bar{k}_2)}
\\ [12pt]
& \quad \exp(\bar{z}(-\bar{k}_1 + \bar{k}_2 - \bar{k}_3) - z(-k_1 + k_2 -
k_3))\ \dst\frac{d^2k_1 d^2k_2 d^2k_3}{\pi^3}
\end{array}
$$

{\bf Lemma}.
\begin{em}
Consider the following functionals of $\{ \alpha(k, \bar{k}) \}$ :
\begin{equation}
\begin{array}{l}
\tilde{\mu}_1 (k, \bar{k}, z, \bar{z}) = 1 + \dst\sum^{\infty}_{n =
1}
\\ [12pt]
\quad \dst\int \frac{\alpha(k_1, \bar{k}_1) \bar{\alpha}(k_2,
\bar{k}_2) \ldots \alpha(k_{2n - 1}, k_{2n - 1}) \bar{\alpha}(k_{2n},
k_{2n})}{(\bar{k}_2 - \bar{k}_1) (k_3 - k_2) (\bar{k}_4 - \bar{k}_3)
\ldots (k_{2n - 1}, k_{2n - 2}) (\bar{k}_{2n} - \bar{k}_{2n - 1}) (k
- k_{2n})}
\\ [12pt]
\quad \cdot \exp(\bar{z}(-\bar{k}_1 + \bar{k}_2 - \bar{k}_3 + \ldots -
\bar{k}_{2n - 1} + \bar{k}_{2n})
\\ [12pt]
- z(-k_1 + k_2 - k_3 + \ldots - k_{2n - 1} + k_{2n}))
\ \dst\frac{d^2k_1 \ldots d^2k_n}{\pi^{2n}} ,
\end{array}
\end{equation}

\vspace{2pt}

\begin{equation}
\begin{array}{l}
\tilde{\mu}_2 (k, \bar{k}, z, \bar{z}) = \dst\sum^{\infty}_{n =
0}
\\ [12pt]
\quad \dst\int \frac{\bar{\alpha}(k_1, \bar{k}_1) \alpha(k_2,
\bar{k}_2) \ldots \bar{\alpha}(k_{2n - 1}, \bar{k}_{2n - 1})
\alpha(k_{2n}, k_{2n}) \bar{\alpha}(k_{2n + 1}, \bar{k}_{2n +
1})}{(k_2 - k_1) (\bar{k}_3 - \bar{k}_2) (k_4 - k_3)
\ldots (\bar{k}_{2n - 1} - \bar{k}_{2n - 2}) (k_{2n} - k_{2n - 1})
(\bar{k}_{2n + 1} - \bar{k}_{2n}) (k - k_{2n + 1})}
\\ [12pt]
\quad \exp(\bar{z}(\bar{k}_1 - \bar{k}_2 + \ldots +
\bar{k}_{2n + 1}) - z(k_1 - k_2 + k_3 - \ldots +
k_{2n - 1})) ,
\\ [12pt]
\quad \dst\frac{d^2k_1 \ldots d^2k_{2n + 1}}{\pi^{2n + 1}}
\end{array}
\end{equation}

\vspace{2pt}

\begin{equation}
\begin{array}{l}
\tilde{q}(z, \bar{z}) = -2 \dst\sum^{\infty}_{n = 0}
\\ [12pt]
\quad \dst\int \frac{\alpha(k_1, k_1) \bar{\alpha}(k_2,
k_2) \ldots \alpha(k_{2n + 1}, \bar{k}_{2n + 1})}{(k_2 - k_1)
(\bar{k}_3 - \bar{k}_2) (k_4 - k_3) \ldots (\bar{k}_{2n + 1} - \bar{k}_{2n})}
\\ [12pt]
\quad \cdot \exp(\bar{z}(-\bar{k}_1 + \bar{k}_2 - \ldots +
\bar{k}_{2n} - \bar{k}_{2n + 1}) - z(-k_1 + k_2 - \ldots +
k_{2n} - k_{2n + 1}))
\\ [12pt]
\quad \dst\frac{d^2k_1 \ldots d^2k_{2n + 1}}{\pi^{2n
+ 1}} .
\end{array}
\end{equation}

We can substitute in $(4 )$--$(6)$ $\alpha(k, \bar{k})$ as a functional in
$q(z, \bar{z}),$ given by $(1)$. As a result of this substitution we
will obtain functionals in $\{ q(z, \bar{z}) \}$ ,given by formal series.
Moreover,

$$
\begin{array}{l}
\tilde{\mu}_1 (k, \bar{k}, z, \bar{z}) = \tilde{\mu}_1 (k, \bar{k},
z, \bar{z})
\\ [12pt]
\tilde{\mu}_2 (k, \bar{k}, z, \bar{z}) = \tilde{\mu}_2 (k, \bar{k},
z, \bar{z})
\\ [12pt]
\tilde{q}(z, \bar{z}) = q(z, \bar{z})
\end{array}
$$
\end{em}

{\bf Theorem 4.1 ( Solution of the Davey-Stewartson equation-II )}

Consider the DS equation
$$
\begin{array}{l}
i \dst\frac{\partial}{\partial t} q(z, \bar{z}, t) = -(\partial^2 +
\bar{\partial}^2) q(z, \bar{z}, t)
\\ [12pt]
+ \dst\frac{1}{2} q(z, \bar{z}, t) (\bar{\partial}^{-1} \partial +
\partial^{-1} \bar{\partial}) (|q(z, \bar{z}, t)|^2)
\\ [12pt]
q(z, \bar{z}, 0) = q(z, \bar{z})
\end{array}
$$
(here $\partial = \dst\frac{\partial}{\partial z}$, $\bar{\partial} =
\dst\frac{\partial}{\partial \bar{z}}$, $\bar{\partial}^{-1} f(z,
\bar{z}) = \dst\frac{1}{\pi} \int \frac{f(z^1, \bar{z}^1)}{z - z^1}
\ d^2z^1$.

\vspace{4pt}

The solution is
\begin{equation}
\begin{array}{l}
q(z, \bar{z}, t) = -2 \dst\sum^{\infty}_{n = 0}
\\ [12pt]
\dst\int \frac{\alpha(k_1, \bar{k}_1, t) \bar{\alpha}(k_2, k_2, t)
\ldots \alpha(k_{2n + 1}, \bar{k}_{2n + 1}, t)}{(k_1 - k_2)(\bar{k}_2
- \bar{k}_3)(k_3 - k_4) \ldots (\bar{k}_{2n} - \bar{k}_{2n + 1})}
\\ [12pt]
\cdot \exp(\bar{z}(-\bar{k}_1 + \bar{k}_2 - \ldots +
\bar{k}_{2n} - \bar{k}_{2n + 1}) - z(-k_1 + k_2 - \ldots +
k_{2n} - k_{2n + 1}))
\\ [12pt]
\dst\frac{d^2k_1 \ldots d^2k_n}{\pi^{2n + 1}}
\end{array}
\end{equation}
\begin{equation}
\begin{array}{l}
\alpha(k, \bar{k}) = \dst\sum^{\infty}_{n = 0}
\\ [12pt]
\dst\int \frac{q(z_1, \bar{z}_1) \bar{q}(z_2, \bar{z}_2) q(z_3,
\bar{z}_3) \ldots \bar{q}(z_{2n}, \bar{z}_{2n}) q(z_{2n + 1}, \bar{z}_{2n +
1})}{(\bar{z}_1 - \bar{z}_2)(z_2 - z_3) (\bar{z}_3 - \bar{z}_4)
\ldots (z_{2k} - z_{2k + 1})}
\\ [12pt]
\cdot \exp(\bar{k}(-\bar{z}_1 - \bar{z}_2 + \bar{z}_3 - \ldots +
\bar{z}_{2k + 1} - k(z_1 + z_2 + z_3 - \ldots +
z_{2n + 1}))
\\ [12pt]
\dst\frac{d^2z_1 d^2z_2 \ldots d^2z_{2n + 1}}{(2\pi)^{2n + 1}}
\end{array}
\end{equation}

$$
\alpha(k,\bar{k},t) = \alpha(k,\bar{k})e^{i(k^2 + {\bar{k}}^2)t} .
$$

{\bf Proof}.

1) Let us compute $q \bar{\partial}^{-1} \partial (|q|^2)$ for $q(z,
\bar{z}, t)$ given by (6 ):
$$
\begin{array}{l}
q \bar{\partial}^{-1} \partial (|q|^2) = - 8 \dst\sum^{\infty}_{n =
1}
\\ [12pt]
\dst\int \frac{\alpha(k_1, \bar{k}_1, t) \bar{\alpha}(k_2, \bar{k}_2,
t) \ldots \alpha(k_{2n + 1}, \bar{k}_{2n + 1}, t)}{(k_1 - k_2)
(\bar{k}_2 - \bar{k}_3) (k_3 - k_4) \ldots (\bar{k}_{2n} -
\bar{k}_{2n + 1})}
\\ [12pt]
\cdot \exp(\bar{z}(-\bar{k}_1 + \bar{k}_2 - \ldots +
\bar{k}_{2n} - \bar{k}_{2n + 1}) - z(-k_1 + k_2 - \ldots +
k_{2n} - k_{2n + 1}))
\end{array}
$$
$$
\begin{array}{l}
\phantom{
\cdot \exp(\bar{z}(-\bar{k}_1 + \bar{k}_2 - \ldots +
\bar{k}_{2n} - \bar{k}_{2n + 1}) - z(-k_1 + k_2 - \ldots +
k_{2n} - k_{2n + 1}))
}
\\
\cdot \dst\sum_{m_1, m_2 \geq 0 \atop m_1 + m_2 + 1 \leq n} (k_{2m_1
+ 1} - k_{2m_1 + 2}) (\bar{k}_{2m_1 + 2m_2 + 2} - \bar{k}_{2m_1 +
2m_2 + 3})
\\ [18pt]
\dst\frac{k_{2m_1 + 2} - k_{2m_1 + 3} + \ldots - k_{2n +
1}}{\bar{k}_{2m_1 + 2} - \bar{k}_{2m_1 + 3} + \ldots - \bar{k}_{2n + 1}}
\\ [12pt]
\dst\frac{d^2k_1 \ldots d^2k_n}{\pi^{2n + 1}}
\end{array}
$$

In the sum over $m_1, m_2$ let us sum over $m_2$ first, then the
second multiplier and the denumerator cancels:
$$
\begin{array}{l}
\dst\sum_{m_1, m_2 \geq 0 \atop m_1 + m_2 + 1 \leq n} (k_{2m_1 + 1} -
k_{2m_1 + 2})(\bar{k}_{2m_1 + 2m_2 + 2} - \bar{k}_{2m_1 + 2m_2 + 3})
\\ [18pt]
\dst\frac{k_{2m_1 + 2} - k_{2m_1 + 3} + \ldots - k_{2n +
1}}{\bar{k}_{2m_1 + 2} - \bar{k}_{2m_1 + 3} + \ldots
- \bar{k}_{2n + 1}}
\\ [12pt]
= \dst\sum^{n - 1}_{m_1 \geq 0} (k_{2m_1 + 1} - k_{2m_1 + 2})
(\bar{k}_{2m_1 + 2} - \bar{k}_{2m_1 + 3} \bar{k}_{2m_1 + 4} -
\bar{k}_{2m_1 + 5} + \ldots + \bar{k}_{2n} - \bar{k}_{2n + 1})
\\ [12pt]
\cdot \dst\frac{(k_{2m_1 + 2} - k_{2m_1 + 3} + \ldots - k_{2n +
1})}{\bar{k}_{2m_1 + 2} - \bar{k}_{2m_1 + 3} + \ldots - \bar{k}_{2n +
1}}
\\ [12pt]
= \dst\sum^{n - 1}_{m_1 = 0} (k_{2m_1 + 1} - k_{2m_1 + 2}) (k_{2m_1 +
2} - k_{2m_1 + 3} + \ldots - k_{2n + 1})
\\ [12pt]
= \dst\frac{1}{2} \left( -(k_1 - k_2 + k_3 - \ldots + k_{2n + 1})^2 -
\dst\sum^n_{p = 1} {k_{2p}}^2 + \dst\sum^n_{p = 0} {k_{2p + 1}}^2 \right)
\end{array}
$$
Let us substitute $q(z, \bar{z}, t)$ as a functional of $\{ \alpha(k,
\bar{k}, t)\}$ in the equation
$$
\begin{array}{l}
\dst\int (i \partial_t + {k_1}^2 + {\bar{k}_1}^2) \alpha(k_1, k_1, t)
\exp(-\bar{z}_1, \bar{k}_1 + z_1 k_1) \frac{d^2k_1}{\pi}
\\ [12pt]
+ \dst\sum^{\infty}_{n = 1} \int \left( i \partial_t + \dst\sum^n_{p
= 0} (k_{2p+ 1}^2 + \bar{k}_{2p + 1}^2) - \sum^n_{p = 1} ({k_{2p}}^2
+ {\bar{k}_{2p}}^2) \right)
\\ [12pt]
\dst\frac{\alpha(k, \bar{k}, t) \bar{\alpha}(k_2, \bar{k}_2, t)
\ldots \alpha(k_{2n + 1}, \bar{k}_{2n + 1}, t)}{(k_1 - k_2)(\bar{k}_2
- \bar{k}_3)(k_3 - k_4) \ldots (\bar{k}_{2n} - \bar{k}_{2n + 1})}
\\ [12pt]
\exp(\bar{z}(-\bar{k}_1 + \bar{k}_2 - \ldots + \bar{k}_{2n} -
\bar{k}_{2n + 1}) - z(-k_1 + k_2 - \ldots + k_{2n} - k_{2n + 1}))
\\ [12pt]
\dst\frac{d^2k_1 \ldots d^2k_{2n + 1}}{\pi^{2n + 1}} = 0
\end{array}
$$
if $\alpha(k, \bar{k}, t) = \alpha(k, \bar{k}) e^{i(k^2 + \bar{k}^2)
t}$

2) Substitution of $\alpha(k, \bar{k}, 0) = \alpha(k, \bar{k})$,
where $\alpha(k, \bar{k})$ is given by (8 ), into (7 ) gives $q(z,
\bar{z}, 0) = q(z, \bar{z})$.

\end{document}